\documentclass[twocolumn,tighten]{aastex62}

\usepackage{graphicx}
\usepackage{grffile}
\usepackage{natbib}
\usepackage{amsmath}
\usepackage{hhline}
\usepackage{url}
\usepackage{etoolbox}
\appto\UrlBreaks{\do\-}
\usepackage{enumitem}
\usepackage{tabularx,tabulary}
\usepackage{multirow}
\newcommand{\wm}{W\,m$^{-2}$\;}
\newcommand{\wmn}{W\,m$^{-2}$}
\newcommand{\gkg}{g\,kg$^{-1}$\;}
\newcommand{\gkgn}{g\,kg$^{-1}$}
\newcommand{\so}{S$_{0}$\;}
\newcommand{\son}{S$_{0}$}
\newcommand{\cod}{CO$_{2}$\;}
\newcommand{\codn}{CO$_{2}$}
\newcommand{\oz}{O$_{3}$\;}

\newcommand{\ag}{H$_{2}$O}
\newcommand{\hmr}{q$_{r}$}

\newcommand{\natgeo}{NatGe}
\newcommand{\natco}{NatCo}

\newcommand{\jas}{JAtS}
\newcommand{\ana}{AN}
\newcommand{\asbio}{AsBio}
\renewcommand{\icarus}{Icar}
\renewcommand{\apj}{ApJ}
\renewcommand{\apjl}{ApJL}
\renewcommand{\jgr}{JGR}
\newcommand{\jgrd}{JGRD}
\newcommand{\cldy}{ClDy}
\renewcommand{\grl}{GeoRL}
\newcommand{\pnas}{PNAS}
\newcommand{\ato}{AtO}
\newcommand{\jcli}{JCli}

\newcommand{\five}{NG5~}
\newcommand{\nine}{NG9~}

\accepted{for publication in ApJ}

\begin{document}
	
	\title{Climate Sensitivity to Carbon Dioxide and Moist Greenhouse threshold of Earth-like planets under an increasing solar forcing}
\shorttitle{\cod and Habitability on Earth-like planets}
\shortauthors{Gomez-Leal et al.}
	\author{Illeana Gomez-Leal}
	\affiliation{Cornell Center for Astrophysics \& Planetary Science, 304 Space Sciences Bldg., Cornell University Ithaca, NY14853-6801, USA}
	\affiliation{Carl Sagan Institute, 304 Space Sciences Bldg., Cornell University, Ithaca, NY14853-6801, USA}
	\author{Lisa Kaltenegger}
	\affiliation{Cornell Center for Astrophysics \& Planetary Science, 304 Space Sciences Bldg., Cornell University Ithaca, NY14853-6801, USA}
	\affiliation{Carl Sagan Institute, 304 Space Sciences Bldg., Cornell University, Ithaca, NY14853-6801, USA}
	\author{Valerio Lucarini}
	\affiliation{Department of Mathematics and Statistic, Whiteknights, PO Box 220, University of Reading, Reading RG6 6AX, UK}
	\author{Frank Lunkeit}
	\affiliation{CEN, Institute of Meteorology, University of Hamburg Grindelberg 5, Hamburg D-20144, Germany}
	
	\email{illeana@astro.cornell.edu; gomezleal.gaia@gmail.com}
	
	\begin{abstract}
		\makeatletter{}Carbon dioxide is one of the major contributors to the radiative forcing, increasing both the temperature and the humidity of Earth's atmosphere. If the stellar irradiance increases and water becomes abundant in the stratosphere of an Earth-like planet, it will be dissociated and the resultant hydrogen will escape from the atmosphere. This state is called the moist greenhouse threshold (MGT). Using a global climate model (GCM) of intermediate complexity, we explore how to identify this state for different \cod concentrations and including the radiative effect of atmospheric ozone for the first time. We show that the moist greenhouse threshold correlates with the inflection point in the water vapor mixing ratio in the stratosphere and a peak in the climate sensitivity. For \cod concentrations between 560~ppm and 200~ppm, the moist greenhouse threshold is reached at a surface temperature of 320~K. Despite the higher simplicity of our model, our results are consistent with similar simulations without ozone by complex GCMs, suggesting that they are robust indicators of the MGT. We discuss the implications for inner edge of the habitable zone as well as the water loss timescales for Earth analog planets.

	\end{abstract}
	
	\keywords{
		Planetary Systems:~Earth -- planets and satellites:~terrestrial planets --~atmospheres}
	
	\NewPageAfterKeywords

\section{Introduction}
\label{sec:introduction}
\makeatletter{}The temperature of our planet depends on the solar luminosity, which will increase by 10\% in the next billion years and by 80\% at the end of the Main Sequence \citep[e.g.][]{Gough1981, Bahcall2001, Schroeder2008}. The rising temperatures will increase the evaporation and are expected to cause the loss of the water reserves of our planet \citep[e.g.][]{Kasting1984, Kasting1988, Kasting1993, Kopparapu2013}.  Two main scenarios have been proposed for the water loss: a) if the moist greenhouse effect dominates the climate, the water vapor becomes abundant in the stratosphere, it will be dissociated by the solar UV radiation, and the resultant hydrogen will escape gradually from the planet's atmosphere \citep{Towe1981}; b) if the runaway greenhouse effect dominates, the oceans evaporate, leading to a steamy atmosphere. The surface temperature
will rise above 1800 K \citep{Kopparapu2013} and the water will be rapidly lost.

The moist greenhouse state and the runaway greenhouse state are used to define the inner boundary of the conservative Habitable Zone \citep[e.g.][]{Kasting1988, Kasting1993, Kopparapu2013, Ramirez2014, Ramirez2016}. The study of the radiative conditions that lead to them is essential to understand the evolution of our planet's climate, as well as the habitability of exoplanets  \citep[e.g.][]{Abe2011, Wordsworth2013, Yang2014}. The planetary climate changes according to the amount of the solar irradiance. As the temperature rises and the humidity is enhanced, the opacity of the atmosphere increases, limiting the outgoing longwave radiation (OLR) of the planet. The OLR is maximum at the Simpson-Nakajima limit \citep{Simpson1927, Nakajima1992, Goldblatt2013}, when the lower atmosphere becomes opaque to the thermal radiation due to the abundance of water vapor. Thus, at this stage, an increase in the solar irradiance does not produce an increase in the emission, but a large global warming, raising the temperatures further. 1D simulations predict that the Earth will enter a moist greenhouse state when the total solar irradiance (TSI) will be 1.015 times greater than the present solar constant (\son=1361.27~\wmn) \citep{Kopparapu2013}. This places the moist greenhouse limit at an orbital distance of 0.99~au in our Solar System, very close to Earth's orbit. These same simulations predict that the runaway greenhouse will dominate when the TSI=1.06~\son, which corresponds to 0.97~au. However, these models do not self-consistently simulate the changes in the surface albedo, the distribution of the humidity and the clouds, the variation in the circulation of the atmosphere, and in general many positive and negative feedbacks that play an important role in the climate of the planet. Global climate models (GCMs) consider these factors, but the simulation of the hydrological cycle and the clouds is far from a trivial task, especially in exotic conditions. Recent 3D studies have shown discrepancies in the concentration of water vapor, the cloud cover, and the evolution of the planetary albedo with an increasing solar radiation \citep{Leconte2013, Wolf2015}. These studies use 1D standards to identify the moist greenhouse threshold in their data: a saturated troposphere and a water vapor mixing ratio of \hmr=3~\gkg  \citep{Kasting1993}. \cite{Leconte2013}, using the LMD Generic GCM (LMDG), did not find a moist greenhouse state, but a runaway state since the troposphere remained unsaturated and the water vapor mixing ratio had a lower value than 1D models. \cite{Wolf2015}, using the Community Atmosphere Model version 4 (CAM4), found two possible moist greenhouse states: one at TSI=1.125~\so and \hmr=1.5\,10$^{5}$~\gkgn, correlating with an abrupt increase in the climate sensitivity; the other, at TSI=1.190~\son, was identified following the 1D standard value of \hmr=3~\gkgn. Both models showed that the troposphere is unsaturated at these states, contrary to 1D models predictions. More recently, \cite{Kasting2015} has obtained a water vapor mixing ratio of 10$^{-6}$~\gkg at the moist greenhouse threshold, using an improved 1D radiative transfer. These studies have shown that the value of the water mixing ratio in the stratosphere depends heavily on the model used. In addition, these models did not include ozone, therefore the temperature and humidity of the atmosphere at the present solar irradiance differed from those on Earth.

Carbon dioxide contributes to 60\% of the global radiative forcing of Earth's atmosphere \citep[e.g.][]{Hartmann2013}, raising the temperature and enhancing the
hydrologic cycle \citep{Manabe1975}. The radiation forcing due to a variation in the \cod concentration between a reference state [\codn]$_{r}$ and the state of interest [\codn]$_{x}$ follows the logarithmic dependence
 \begin{equation}\label{arr}
\Delta F_{x,r}=5.35 \ln\Bigg( \frac{[CO_{2}]^{r}}{[CO_{2}]^{x}}\Bigg) 
\end{equation}
where $\Delta F_{x,r}$ is in \wm \citep[e.g.][]{Arrhenius1896, Myhre1998}. According to this relation, a doubling of the \cod concentration is estimated to produce a radiative forcing of 3.7~\wm by the Intergovernmental Panel on Climate Change (IPCC) assessments \citep{Collins2013}, as well as by radiative transfer models \cite[e.g.][]{Etminan2016}, and by global climate models \citep{Myhre2013}. The solar irradiance required for the transition from a snowball state to a warm state of an Earth-like planet decreases with increasing atmospheric \cod \citep{Boschi2013}. Therefore, the \cod concentration in the atmosphere may have an influence in the planetary habitability.

The greenhouse effect of ozone on Earth's climate is also important. Ozone absorbs most of the solar UV radiation through photodissociation, warming the stratosphere.  This produces a temperature inversion that determines the level of the tropopause, and increases the temperature and the humidity of the lower atmosphere, as well as the surface temperature. In the last decades, ozone has contributed with about 0.35~\wm, due to its increase in the troposphere by human activities \citep{Forster2007}, and about -0.05~\wmn, due to its decrease in the stratosphere, which is equivalent to 20\% of the \cod contribution. Including atmospheric ozone is essential to describe Earth's current climate. In the absence of ozone, the temperature inversion in the stratosphere does not occur, UV radiation penetrates deeply in the atmosphere, the water dissociation rates increase, and the temperature gradient extends to higher levels. 

The amount of water vapor is enhanced at higher temperatures, having both chemical and radiative effects on the atmosphere. The products of the dissociation of water such as HO$_{x}$ radicals increase, depleting ozone. However, they also increase HNO$_{3}$ and remove NO$_{2}$, slowing the ozone depletion. In addition, water vapor absorbs latent heat, cooling the environment and decreasing the reaction rates. Several studies have shown that the radiative effect dominates and increasing water vapor in the atmosphere only depletes ozone in the tropical lower stratosphere and the high latitudes of the southern hemisphere, while elsewhere ozone increases \citep[e.g.][]{Evans1998, Tian2009}. This indicates that the radiative effect of ozone is enhanced in a warmer atmosphere. However, the interaction between water and ozone is complex to simulate and it has not been studied at high solar irradiances with 3D models.

We use the intermediate complexity model Planet Simulator (PlaSim) to explore the climate sensitivity of an Earth-like planet, including \oz and different \cod concentrations, as a function of the solar irradiance, in order to constrain the characteristics of the moist greenhouse threshold and to derive a new reference quantity for this state. The article is organized as follows: In Section 2, we describe the model and the methods. In Section 3, we present our results. First, we validate our model for the present solar irradiance (\son=1361.27~\wmn), by simulating Earth's climate with the present \oz concentration and 388~ppm of \cod, and we compare our results with atmospheric reanalysis data and previous studies using satellite observations. We test the response of our model to solar forcing by doubling the \cod pre-industrial concentration (560~ppm) and we compare it to established values. Then, we study the contribution of carbon dioxide to the greenhouse effect, simulating the present Earth with four different \cod concentrations: 388~ppm (taken as representative of present conditions), 280~ppm (a pre-industrial level), 200~ppm (a typical value during Earth glaciations) and 560~ppm (the double of the pre-industrial concentration). We increase the solar irradiance in subsequent simulations until the atmosphere becomes entirely opaque. As a first approach, we include \oz in a concentration equal to the present values. We study the variation of the surface temperature, the surface albedo, the cloud radiative effect, the Bond albedo, the water vapor mixing ratio, the stratospheric temperature, and the emissivity of the atmosphere. Finally, we estimate the moist greenhouse threshold and the water lifetime of an Earth analog and we compare our results to previous GCM studies. In Section 4, we discuss the results and we present our conclusions in Section 5.

\section{Model and Methods}\label{mod}
We have used the intermediate complexity model Planet Simulator (PlaSim)\footnote{freely available at https://www.mi.uni-hamburg.de/en/ arbeitsgruppen/theoretische-meteorologie/modelle/plasim.html} \citep{Fraedrich2005a, Fraedrich2005b, Lunkeit2011} to simulate the global warming of the Earth under an increasing solar irradiance for several \cod concentrations. While being simpler than the state-of-the-art global climate models in terms of resolution and adopted parameterizations, intermediate complexity models represent a compromise between sophistication and computation time. PlaSim can simulate a large variety of scenarios and allows us to examine aspects of the climate in a very efficient manner, performing a large number of simulations in a short time. As a result, the model has been instrumental in studying climate change using rigorous methods of statistical mechanics \citep{Ragone2016}. It has the advantage of a great degree of flexibility and robustness when terrestrial, astronomical, and astrophysical parameters are altered. It has been extensively used for studying climate sensitivity to the variation of solar radiation \citep{Lucarini2010a,Lucarini2010b, Lucarini2013}, \cod concentration \citep{Boschi2013}, obliquity \citep{Kilic2017}, eccentricity \citep{Linsenmeier2015}, and ozone \citep{Bordi2012}.

The primitive equations for vorticity, divergence, temperature, and surface pressure are solved via the spectral transform method \citep{Eliasen1970, Orszag1970}. The parameterization in the shortwave (SW) radiation follows \cite{Lacis1974} for the cloud free atmosphere. Transmissivities and albedos for high, middle, and low level clouds are parameterized following \cite{Stephens1978} and \cite{Stephens1984}. The downward radiation flux density $F^{\downarrow SW}$ is the product of different transmission factors with the solar flux density ($\mathcal{E}_{0}$) and the cosine of the solar zenith angle ($\mu_{0}$) as
\begin{equation}
F^{\downarrow SW}=\mu _{0}\mathcal{E}_{0}\cdot \mathcal{T}_{R}\cdot \mathcal{T}_{H_{2}O}\cdot \mathcal{T}_{O_{3}}\cdot \mathcal{T}_{C}\cdot \mathcal{R}_{S}
\end{equation}
which includes the transmissivities due to Rayleigh scattering ($R$), and cloud droplets ($C$), water (H$_{2}$O) and ozone (O$_{3}$) absorption (Chappuis band), and $R_{S}$ comprises different surface albedo values. $\mathcal{E}_{0}$ and $\mu_{0}$ are computed following \cite{Berger1978a, Berger1978b}. For the clear sky longwave (LW) radiation, the broad band emissivity method is employed \citep{Manabe1961, Rodgers1967, Sasamori1968, Katayama1972, Boer1984}.
\begin{equation}
F^{\uparrow LW}(z)= \mathcal{A}_{S}B(T_{S}) \mathcal{T}_{(z,0)}+\int_0^z B(T')\frac{\delta \mathcal{T}_{(z,z')}}{\delta z'}
\end{equation}
\begin{equation}
F^{\downarrow LW}(z)= \int_\infty^z B(T')\frac{\delta \mathcal{T}_{(z,z')}}{\delta z'}
\end{equation}
where $B(T)$ denotes the blackbody flux and $\mathcal{A_{S}}$ is the surface emissivity. The transmissivities for water vapor, carbon dioxide, and ozone are taken from \cite{Sasamori1968}. These empirical formulas are obtained from meteorological data and are dependent on the effective amount of each gas. The effective amount is obtained as
\begin{equation}
u_{X}(p,p')=\frac{1}{g}\int_{p}^{p'}q_{X}(\frac{p''}{p_{0}})dp''
\end{equation}
where $g$ is the gravitational acceleration, $q_{X}$ is the mixing ratio, $p$ is the pressure, $p_{0}$=1000~hPa is the reference pressure.

The \ag\;continuum absorption is parameterized by 
\begin{equation}
\tau _{cont}^{H_{2}O}=1.- exp\,(-0.03\;u_{H_{2}O})
\end{equation}
To account for the overlap between the water vapor and the carbon dioxide bands near 15~\micron, the \cod absorption is corrected by a \ag transmission at 15~\micron given by

\begin{equation}
\mathcal{T}^{15\micron}_{H_{2}O}=1.33 - 0.832\,(u_{H_{2}O}+ 0.0286)^{0.26}
\end{equation}

Cloud flux emissivities are obtained from the cloud liquid water content \citep{Stephens1984} by
\begin{equation}
\mathcal{A}^{cl}=1.- exp(-\beta_{d}k^{cl}W_{L})
\end{equation}
where $\beta_{d}=1.66$ is the diffusivity factor, $k^{cl}$ is the mass absorption coefficient, set to a default value of 0.1~m$^{2}$\,g$^{-1}$ \citep{Slingo1991}, and W$_{L}$ is the cloud liquid water path. For a single layer between z and z' with the fractional cloud cover $C$, the total transmissivity is

\begin{equation}
\mathcal{T}^{*}_{(z,z')}=\mathcal{T}_{(z,z')}(1.- C\mathcal{A}^{cl})
\end{equation}
where $\mathcal{T}_{(z,z')}$ is the clear sky transmissivity. Random overlapping of clouds is assumed for multilayers and the total transmissivity becomes
\begin{equation}
\mathcal{T}^{*}_{(z,z')}=\mathcal{T}_{(z,z')}\Pi_{j}(1.- C_{j}\mathcal{A}^{cl}_{j})
\end{equation}
where $j$ denotes each cloud layer.

It includes dry convection, large-scale precipitation, boundary-layer fluxes of latent and sensible heat, and vertical and horizontal diffusion \citep{Louis1979, Laursen1989, Roeckner1992}. Penetrative cumulus convection is simulated by a moist convergence scheme \citep{Kuo1965, Kuo1974} including some improvements: cumulus clouds are assumed to exist only if the environmental air temperature and moisture are unstable stratified with respect to the rising cloud parcel, and the net ascension is positive. Shallow convection is represented following \cite{Tiedtke1988} and clouds originated by extratropical fronts are simulated considering the moisture contribution between the lifting level and the top of the cloud, instead of the total column. The effects of water, carbon dioxide, and ozone are taken in account in the radiative transfer. However, the interaction between water and ozone depends greatly on temperature, it is still complex to simulate, and therefore, the evolution of the ozone concentration with an increasing solar irradiance is not yet well understood \citep{Evans1998, Tian2009}. As a first approach, we use the present distribution in all our simulations, in order to take in account its radiative effect as a function of the solar irradiance. The ozone concentration is prescribed following the distribution described by \citet{Green1964},
\begin{equation}\label{ozo}
u_{O_{3}}(z)=(\alpha+\alpha e^{-\beta/c})/(1+e^{(z-\beta)/c})
\end{equation} 
where $u_{O_{3}}(z)$ is the ozone concentration in a vertical column above the altitude $z$, $\alpha$ is the total ozone in the vertical column above the ground, $\beta$ is the altitude where the ozone concentration is maximal, and $c$ is a fitting parameter. Equation~\ref{ozo} fits closely the mid-latitude winter ozone distribution with $\alpha= 0.4$~cm, $\beta= 20$~km, and $c = 5$~km. The latitudinal variation and the annual cycle are modeled by introducing the dependence of $\alpha$ with time,
\begin{equation}
\alpha(t,\phi)=\alpha_{0}+\alpha_{1}\,|sin\,\phi|\,+\alpha_{c}\,|sin\,\phi|\, cos(2\pi(d-d_{off})/n) 
\end{equation}
where $t$ is time, $\phi$ is latitude, $d$ is the day of the year, d$_{off}$ is an offset, $n$ is the number of days per year, $\alpha_{0}$=0.25, $\alpha_{1}$=0.11, and $\alpha_{c}$=0.08. This representation of the ozone distribution is simple, but it allows studying the sensitivity of the problem. The global atmospheric energy balance is improved by re-feeding the kinetic energy losses due to surface friction and horizontal and vertical momentum diffusion \citep{Lucarini2010b}.  A diagnostic of the entropy budget is available \citep{Fraedrich2008}. The average energy bias on the energy budget is smaller than 0.5~\wm in all simulations, which it is achieved locally by an instantaneous heating of the air \citep{Lucarini2011}.

We have used a T21 horizontal resolution (5.6$^{\circ}$$\times$5.6$^{\circ}$ on a gaussian grid) and 18 vertical levels with the uppermost level at 40~hPa. This resolution enables to have an accurate representation of the large scale circulation features and the global thermodynamical properties of the planet \citep{Pascale2011}. Our simulations include a 50 m mixed-layer ocean and a thermodynamic
sea-ice model. The surface energy budget has been calculated as,
\begin{equation}
\Delta E= F^{net}_{SW}-F^{net}_{LW}-F^{net}_{LH}-F^{net}_{SH}-\rho_{w}L_{f}v_{SM}
\end{equation}
where $F^{net}_{SW}$ is the net shortwave radiative flux, $F^{net}_{LW}$ is the net longwave radiative flux, $F^{net}_{LH}$ is the latent heat flux, $F^{net}_{SH}$ is the sensible heat flux, $\rho_{w}$ is the density of water, $L_{f}$ is the latent heat of fusion, and $v_{SM}$ is the snow melt. The surface is in equilibrium at every state, with an energy budget $<$0.02~\wmn.  We have studied the evolution of the climate increasing the solar irradiance for five values of the \cod atmospheric concentration: 388~ppm (present value), 280~ppm (a pre-industrial level), 200~ppm (a typical value during Earth glaciations) and 560~ppm (double of the pre-industrial level). We increase the solar irradiance from the present value (1361.27~\wmn) until the efficient emissivity of the atmosphere is near unity. Each simulation has a length of 100 years to ensure that the system achieves the equilibrium well before the end of the run and the statistical results are averaged over the last 30 years in order to rule out the presence of transient effects. 

The TSI and the concentrations of \cod and \oz are inputs in the model. The surface temperature (T$_{S}$) is calculated as the global mean of the near surface air temperature. The effective temperature is calculated as the global mean of the radiative temperature at the top-of-the-atmosphere (TOA), $T_{eff}=(F^{TOA}_{LW}/\sigma)^{1/4}$, where F$^{TOA}_{LW}$ is the outgoing longwave radiation at TOA and $\sigma$ is the Stefan-Boltzmann constant. The Bond albedo is calculated as $A=1-(4\, F^{TOA}_{SW}/S_{0})$, where F$^{TOA}_{SW}$ is the reflected radiation at TOA and \so is the solar constant. The normalized greenhouse parameter is calculated as $g_{n}=1-(T_{eff}/T_{S})^{4}$. The CRE is calculated as the difference between the upward flux for clear-sky and for all-sky conditions, $CRE=F^{up}_{clear-sky}-F^{up}_{all-sky}$, for both SW and LW ranges. The global mean temperature (T$_{40}$) and the water vapor mixing ratio (q$_{r}$) in the stratosphere are calculated at 40~hPa. This level corresponds to an altitude about 25~km on the present conditions, and it represents a compromise between the concentration and the dissociation of O$_{2}$, \ag, and \oz \citep[e.g.][]{Garcia1983, Fioletov2008}. The standard deviation is at least one order of magnitude lower than the values of the results presented in Tables~\ref{tab2} and \ref{tab3},thus, the errorbars in the figures are too small to be seen.

Water is a trace gas (volumetric concentration $\chi\leq1$\%) in the present atmosphere, but at higher temperatures it may become dominant, having an impact on the variation of the atmospheric mass and pressure (non-diluted regime). We calculate the water mixing ratio corresponding to $\chi=1$\% as $q_{t}=m_{w}/m_{\alpha}=\chi\,(M_{w}/M_{\alpha})=6.3$~\gkgn, where $m_{w}$ and $m_{a}$ represent the masses of water and air, respectively, and $M_{w}$ and $M_{a}$ are the molar masses. At the MGT, \hmr$\sim7.5$~\gkg at 40~hPa (see Section 3), which corresponds to $\chi\approx1.1$\%. Since \hmr$\sim q_{t}$, water vapor is not dominant in the atmosphere and we have not used the non-diluted regime of water in our simulations.

\subsection{The Moist Greenhouse Threshold}
\citet{Kasting1993} defined the MGT as the state at which the water vapor mixing ratio in the stratosphere increases considerably with the solar forcing, obtaining a value of 3~\gkg using a radiative convective model. This value has
been used by GCM studies to identify the MGT \citep[e.g.][]{Leconte2013, Wolf2015}. 

Instead of using the value calculated by 1D models, we calculate the curve of the water vapor mixing ratio at 40~hPa as a function of solar irradiance using a polynomial approximation for each \cod concentration series. We identify the MGT with the inflection point of the curve, which corresponds to the maximum increase of \hmr\;with solar irradiance, as ($q^{d}_{r}$, S$_{MGT}$), where $q^{d}_{r}$ is the value of the water mixing ratio and S$_{MGT}$ is the TSI at that point. We derive the equivalent orbital distance (D) of  S$_{MGT}$ in the present solar system as $D= (S_{0}/S_{MGT})^{1/2}$, where $D$ is expressed in astronomical units and S$_{MGT}$ is a multiple of the present solar irradiance (\son=1361.27~\wmn). 

Taking in account the bulk effect of the atmosphere, the radiative balance of the planet can be expressed as,
\begin{equation}
\frac{S}{4}~(1-A)=\frac{2-\epsilon}{2}~\sigma T_{S}^{4}
\end{equation}\label{eq2}
where $S$ is the solar irradiance, $A$ is the Bond albedo of the planet, $\epsilon$ is the efficient emissivity of the atmosphere, and $T_{S}$ is the surface temperature. Thus, the efficient emissivity of the atmosphere can be calculated as,
\begin{equation}
\epsilon=2\,(1-(F_{TOA}/F_{S}))
\end{equation}\label{eq3}
where $F_{TOA}$ and $F_{S}$ are the outgoing LW fluxes at the top of the atmosphere and at the surface, respectively. Relevant changes on the climate are identified by the analysis of the climate sensitivity $\zeta$, 
\begin{equation}
\zeta= \frac{\Delta T_{S}}{(1-A)(\Delta S/4)}
\end{equation}\label{eq4}
where $\Delta T_{S}$ is the change in the global surface temperature and $\Delta S$ is the change in the solar irradiance. In order to obtain a better accuracy on the point where these changes occur, we perform a polynomial approximation of the variable series and we derive the inflection point (see Section~\ref{res}).  

\begin{figure}[ht]\centering  
	\includegraphics[width=\columnwidth]{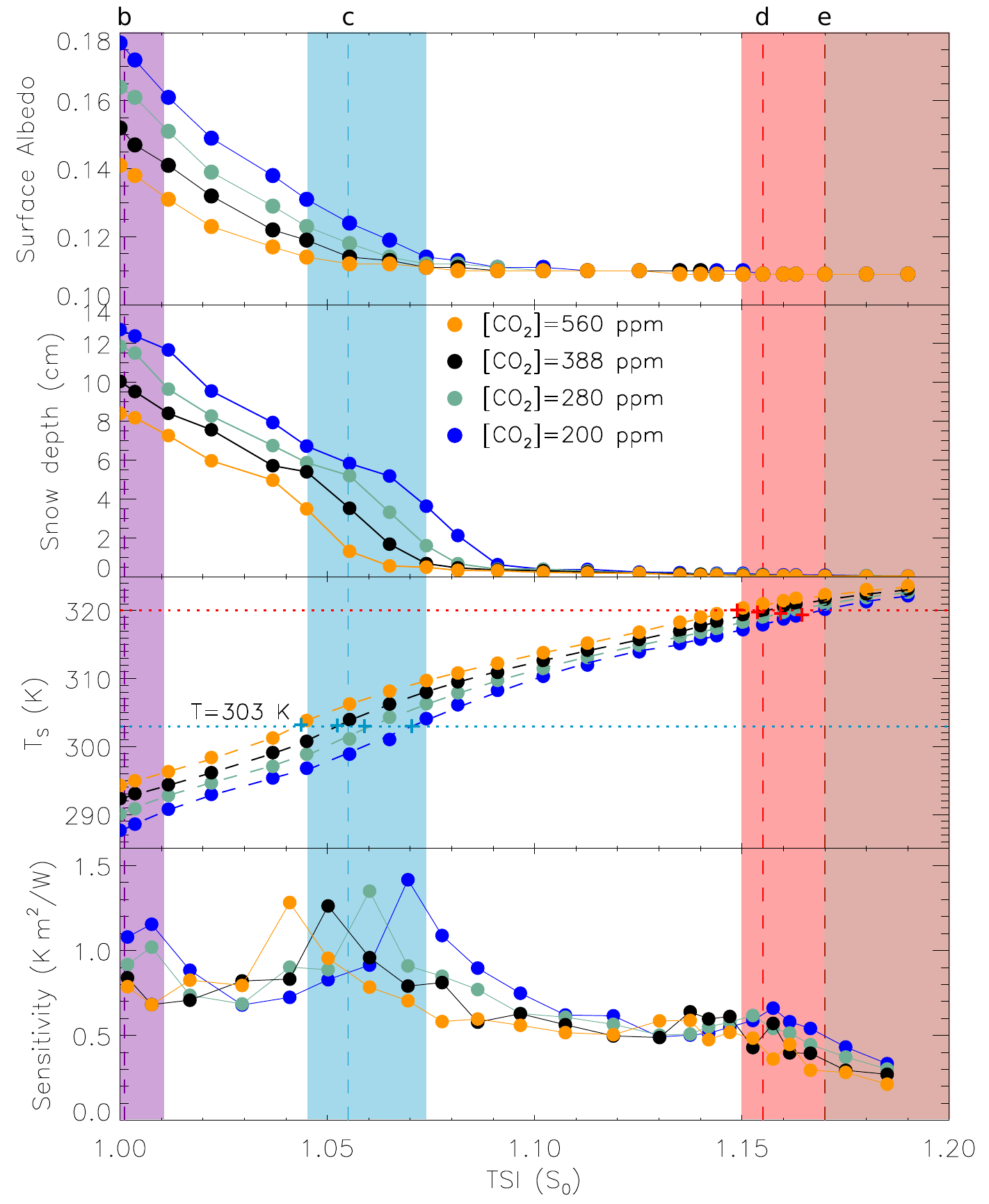}
	\caption{Surface albedo, snow depth, surface temperature, and climate
		sensitivity as a function of the total solar irradiance (TSI, in units of the solar constant, \so) for four \cod concentrations. The dashed lines indicate the states at the maximum cloud fraction (b), the complete melt of the polar ice caps (c), the moist greenhouse initiation (d), and the state where the atmosphere becomes opaque (e) for a [\codn]=388~ppm. The stripes indicate the same state from a [\codn]=560~ppm (left limit) to a [\codn]=200~ppm (right limit). The crosses correspond to the inflection points of the polynomial fittings (dashed curves) of the surface temperature series of each case (Table~\ref{tab3}).\label{Fig1}}
\end{figure}

\begin{figure}[ht]\centering  
	\includegraphics[width=\columnwidth]{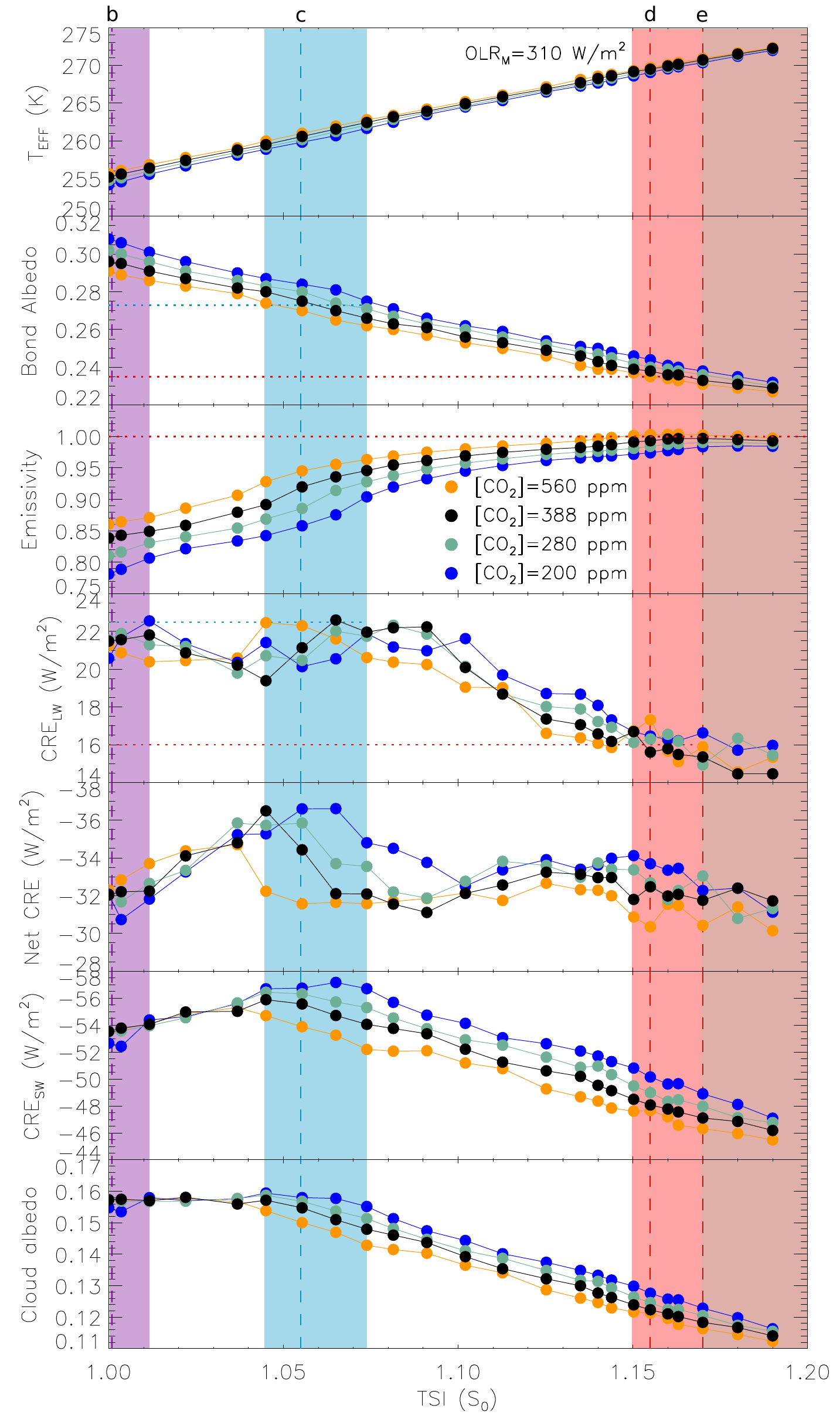}
	\caption{Effective temperature, Bond albedo, emissivity, cloud radiative effect (CRE) for the LW radiation, net CRE, CRE for the SW radiation, and cloud albedo as a function of the total solar irradiance total solar irradiance (TSI) for four \cod concentrations. The dashed lines indicate the states at the maximum cloud fraction (b), the complete melt of the polar ice caps (c), the moist greenhouse initiation (d), and the state where the atmosphere becomes opaque (e) for a [\codn]=388~ppm. The stripes indicate the same state from a [\codn]=560~ppm (left limit) to a [\codn]=200~ppm (right limit). The maximum OLR shown in our simulations is 310~\wmn. \label{Fig2}}
\end{figure}

\begin{figure}[ht]\centering  
	\includegraphics[width=\columnwidth]{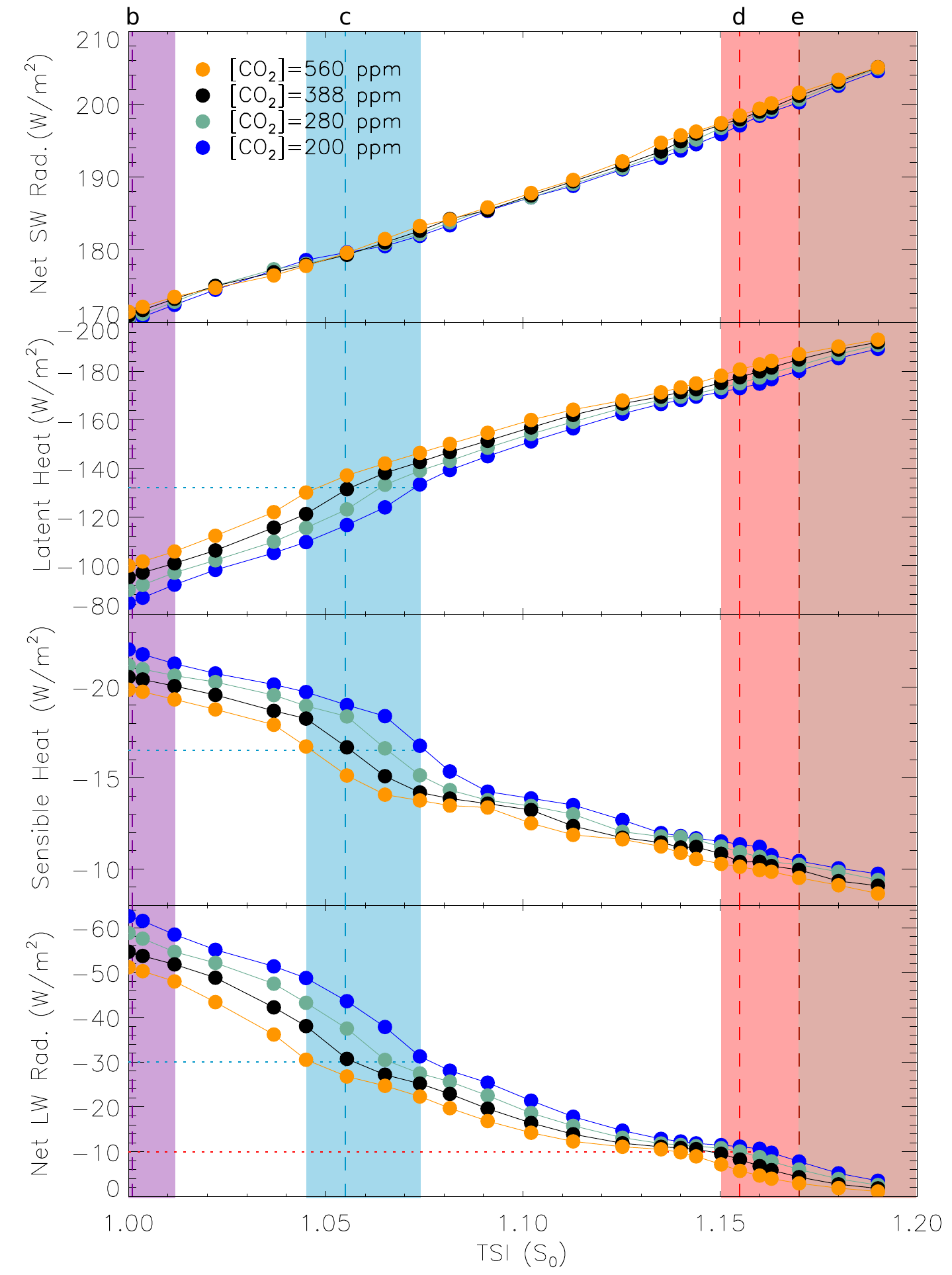}
	\caption{Net SW radiation, latent heat flux, sensible heat flux, and net LW
		radiation (all in \wmn) as a function of the total solar irradiance (TSI, in units
		of the solar constant, \so) for four \cod concentrations. Negative values
		correspond to upward flux. The dashed lines indicate the states at the maximum cloud fraction (b), the complete melt of the polar ice caps (c), the moist greenhouse initiation (d), and the state where the atmosphere becomes opaque (e) for a [\codn]=388~ppm. The stripes indicate the same state from a [\codn]=560~ppm (left limit) to a [\codn]=200~ppm (right limit). The
		horizontal dotted lines indicate a similar value at the same state. \label{Fig3}}
\end{figure}

\begin{figure}[ht]\centering  
	\includegraphics[width=\columnwidth]{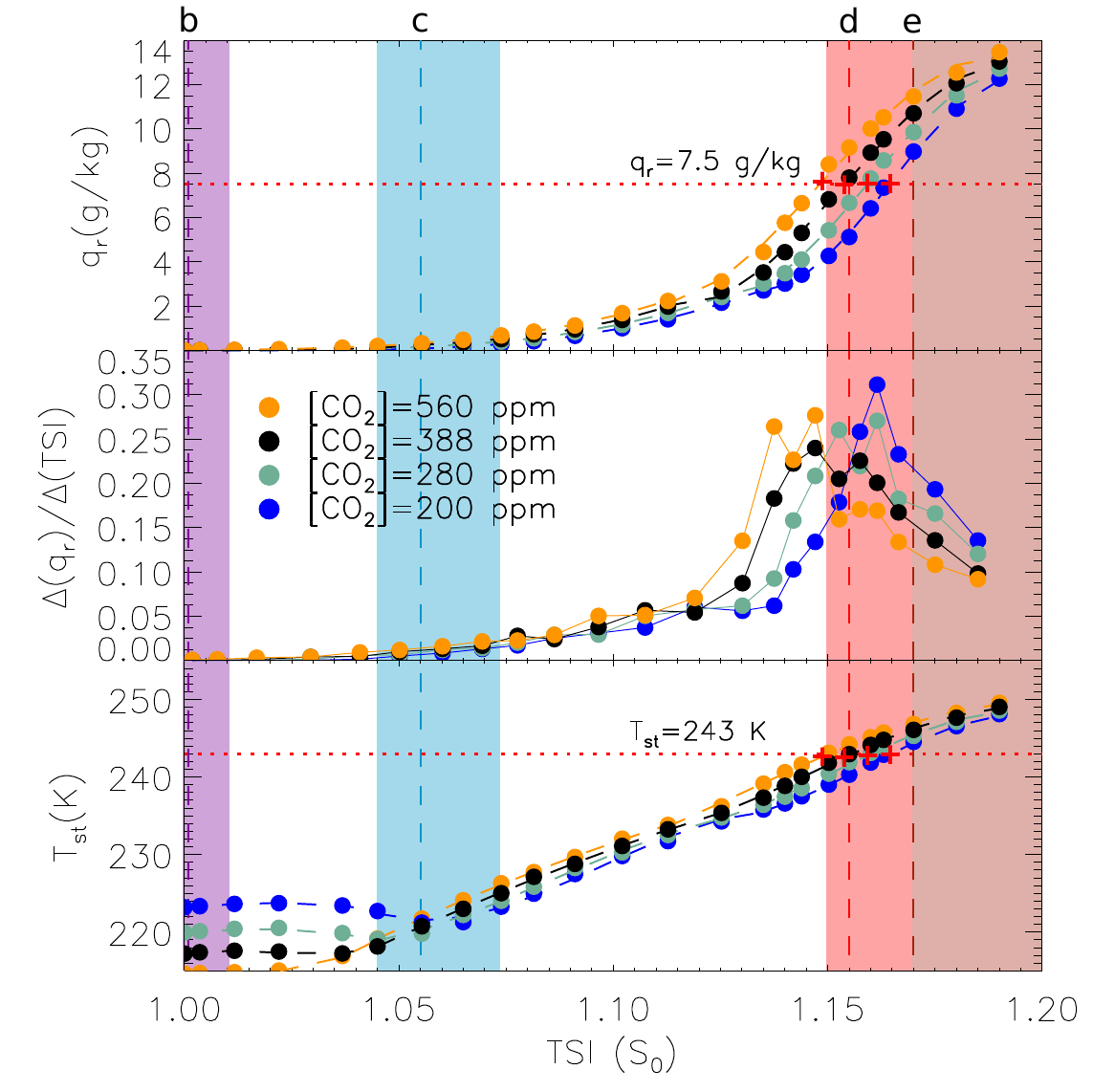}
	\caption{Water vapor mixing ratio (qr), its variation with the solar forcing
		(in $g\,kg^{-1}\,W^{-1}\,m^{2}$), and temperature at 40 hPa as a function of the total solar irradiance (TSI) for four \cod concentrations. The dashed curves are the polynomial fittings of the \hmr, the crosses indicate their inflection points, and the dotted lines their mean value. The vertical dashed lines indicate the states at the maximum cloud fraction (b), the complete melt of the polar ice caps c), the moist greenhouse initiation (d), and the state where the atmosphere becomes opaque (e) for a [\codn]=388~ppm. The stripes indicate the same state from a [\codn]=560~ppm (left limit) to a [\codn]=200~ppm (right limit). \label{Fig4}}
\end{figure}

\subsection{The Water Loss}
In a moist greenhouse state, the water loss is limited by the rate of mass transport to the homopause. Here, we have addressed the simplified problem where water dissociates completely into hydrogen, and despite possible reactions with methane in the stratosphere, the water vapor concentration at the homopause can be approximated to the water vapor concentration at the stratosphere \citep{Hunten1973, Kasting1993, Wolf2015}. These calculations represent a lower limit, since water has to be pumped first into the stratosphere.

We estimate the water loss using the diffusion-limited escape rate of atomic hydrogen, which can be approximated as, 
\begin{equation}
\Phi_{H}\simeq (b/H_{a})\, q_{H}
\end{equation}\label{eq5}
where $b$ is the average binary diffusion coefficient of hydrogen, H$_{a}$ is the scale height, $(b/H_{a})=2.5\times10^{13}$~cm$^{-2}\, s^{-1}$, and q$_{H}$ is the total mixing ratio of hydrogen at the stratosphere ($q_{H}\sim2$\hmr). Taking in account the number of water molecules in the oceans ($n=4.416\times10^{46}$), the total number of hydrogen molecules is $n_{H}=2n$, and the lifetime of water at a certain state can be calculated as,
\begin{equation}
\tau=n_{H}/(a\,\Phi_{H})
\end{equation}\label{eq6}
where $a$ is the global area at 40~hPa, the reference pressure level previously discussed.

\section{Results}\label{res}

\subsection{Climate stages}
The effect of increasing solar irradiance is amplified by a positive ice-albedo feedback: as the surface temperature rises, the snow melts, decreasing the surface albedo and the Bond albedo (Figs.~\ref{Fig1} and \ref{Fig2}), and as a result, the planet absorbs more solar radiation. The latent heat transfer from the surface to the atmosphere aloft is enhanced (Fig.~\ref{Fig3}), the atmosphere becomes more humid and opaque to the thermal radiation, the emissivity and the greenhouse effect increase (Fig.~\ref{Fig2}), and water vapor becomes abundant in the stratosphere (Fig.~\ref{Fig4}). An abrupt variation in the climate sensitivity indicates a change in the overall climate of the planet. Previous GCM studies \citep[e.g.][]{Wolf2015, Popp2016} obtain one climate sensitivity peak corresponding to the moist greenhouse state. PlaSim results show three peaks at different values of the solar irradiance, depending on the concentration of carbon dioxide (Fig.~\ref{Fig1} and Table~\ref{tab1}), indicating important changes in the planet's climate. 

\begin{table}
	\begin{center}
		\centering
		\begin{tabularx}{1.0\columnwidth}{c  c   c   c}
			\cline{1-4}
			[CO$_{2}$] & 1st peak (S$_{0}$) & 2nd peak (S$_{0}$) & 3rd peak (S$_{0}$)\\\hhline{====}
			200 & [1.004, 1.012] & [1.065, 1.074] & [1.155, 1.170] \\
			280 & [1.004, 1.012] & [1.055, 1.065] & [1.150, 1.163] \\
			388 & [1.000, 1.004] & [1.045, 1.055] & [1.144, 1.155] \\
			560 & [1.000, 1.004] & [1.037, 1.045] & [1.135, 1.150] \\\cline{1-4}
		\end{tabularx}
		\caption{Climate sensitivity peaks as a function of the solar irradiance (in units of the solar constant \son) for different \cod concentrations (in ppm). }\label{tab1}
	\end{center}
\end{table} 

Our analysis focus on five climate stages: a) the climate at the present solar irradiance;  b) the first peak of the climate sensitivity, which correlates with an increase in the cloud albedo; c) the second peak of the climate sensitivity, which correlates with the complete melt of the planet’s ice and snow; (d) the third peak, which we identify with the MGT; and (e) the state when the atmosphere becomes opaque.\\

\emph{a) Climate under the present solar irradiance.-- } We compare PlaSim present Earth's climate with the European Centre for Medium-Range Weather Forecasts (ECMWF) climate reanalysis data (ERA)\footnote{\url{http://www.ecmwf.int/en/research/climate-reanalysis/browse-reanalysis-datasets}}, which provides an accurate representation of the current climate of the Earth. Our simulations are at a steady state, contrary to reanalysis data. Nonetheless, since the current climate change is relatively slow, these comparisons are meaningful, identifying any bias of
our model with respect to present Earth conditions. We use ERA-20CM flux data \citep{Hersbach2015} to calculate the global surface temperature, the effective temperature, the Bond albedo of the planet, and the efficient emissivity of the atmosphere for the thermal radiation. The stratospheric temperature and the water mixing ratio have been extracted from ERA-20C data \citep{Poli2013, Poli2016}.

\begin{figure}[ht]\centering  
	\includegraphics[width=\columnwidth]{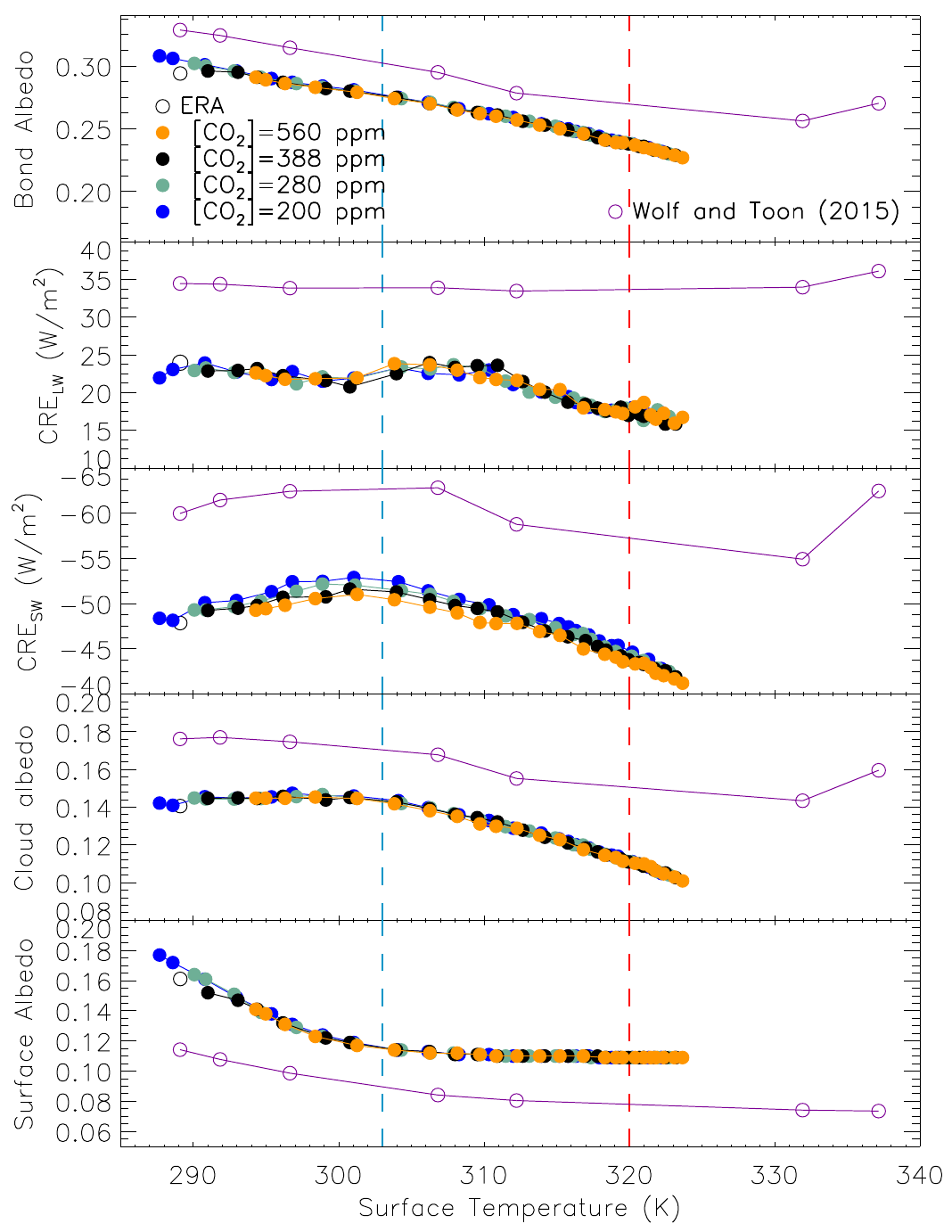}
	\caption{Comparison between the results of this paper, ERA reanalysis data, and \cite{Wolf2015}. From top to bottom: Bond albedo, LW cloud
		radiative effect (CRE), SW CRE, cloud albedo, and surface albedo as function
		of global mean surface temperature. The blue and the red dashed lines
		correspond to the temperature at the complete ice melt and at the moist
		greenhouse threshold, respectively, with a \codn=\,388~ppm. \label{Fig5}}
\end{figure}

\begin{figure*}[ht]\centering  
	\includegraphics[width=\textwidth]{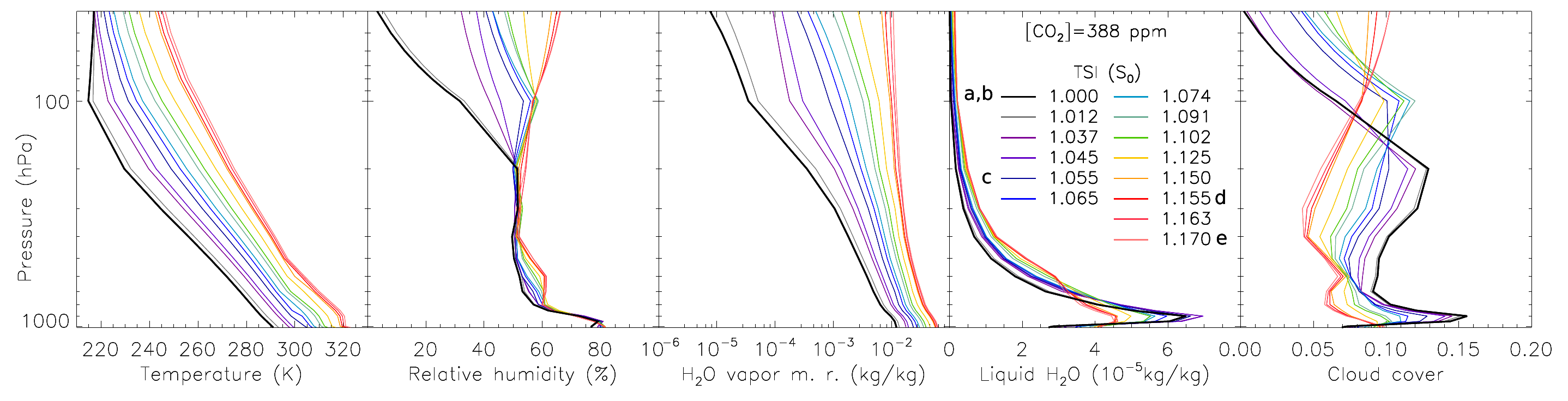}
	\includegraphics[width=\textwidth]{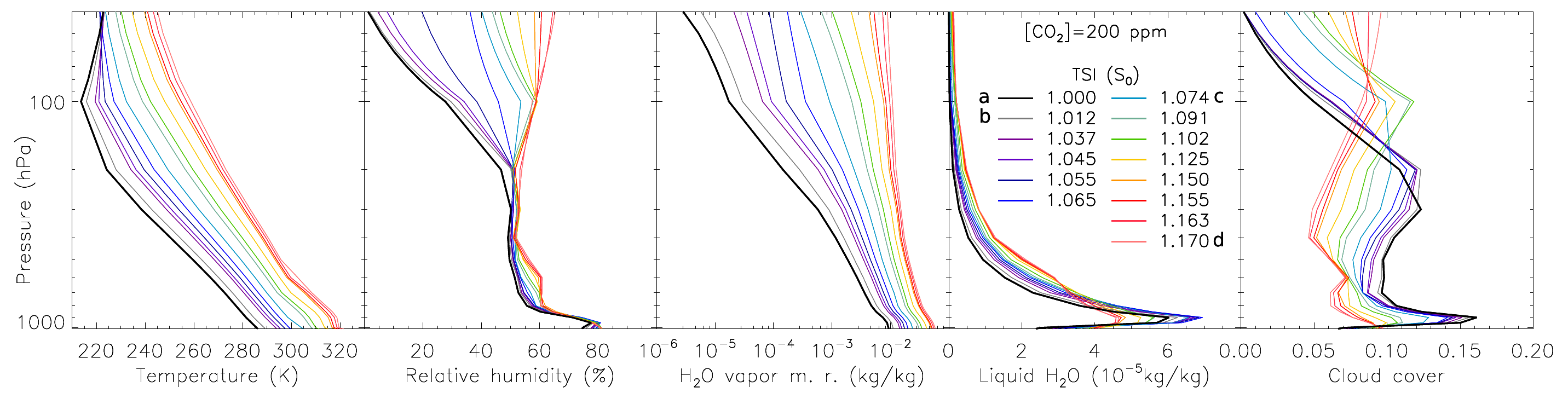}
	\caption{Vertical profiles of surface temperature, relative humidity, water vapor mixing ratio, liquid water content, and cloud cover at several values of the total solar irradiance (TSI, in units of the solar constant \son) for two \cod concentrations. The characters mark the state at the present solar irradiance (a), the state of maximum cloud fraction (b), the state of complete melt of the polar ice caps (c), the state of the moist greenhouse initiation (d), and the state when the atmosphere becomes opaque (e).\label{Fig6}}
\end{figure*}

The surface temperature, the effective temperature, the stratospheric temperature, and the albedo differ by less than 1\% from ERA data at the same \cod concentration (388~ppm) and solar irradiance (TSI=1361.27~\wmn, see Table~\ref{tab2} and Fig.~\ref{Fig5}). The tropopause lies at 200~hPa in ERA and PlaSim, the stratospheric temperature is 216~K and 217~K respectively. Our data show a larger water mixing ratio (7.4$\times$10$^{-3}$~\gkgn) than ERA data (2.3$\times$10$^{-3}$~\gkgn), but the cloud cover and the cloud radiative effect (CRE) are well reproduced (Fig.~\ref{Fig6}). The results for the net solar radiation, latent heat flux, sensible heat flux, and net longwave radiation fluxes are in agreement with satellite measurements \citep{Tren2009}. The surface temperature for the present \cod concentration (388~ppm) is about 2~K higher than for a preindustrial concentration (280~ppm), in agreement with the IPCC reports \citep{Hartmann2013}. The surface albedo changes considerably with \codn, showing a 20\% difference between the present and the pre-industrial value (Fig.~\ref{Fig1}). Larger \cod concentrations show higher surface temperatures and enhanced latent heat fluxes and humidity (Fig~\ref{Fig6}). 

We simulate the response to solar forcing by doubling the \cod concentration (560~ppm) with respect to the pre-industrial level (280~ppm). We obtain an equilibrium climate sensitivity of 2.1~K and a climate feedback parameter of 1.75~W\,m$^{-2}$\,K$^{-1}$, which are within the range of values estimated by the IPCC reports and other recent estimations \citep[e.g.][]{Bindoff2013, Forster2016}.

\emph{b) Increase in the total cloud fraction.-- }\label{b}
The first peak of climate sensitivity coincides with an increase in the cloud fraction (Fig.~\ref{Fig2} and~\ref{Fig6}), which increases the surface temperature. It occurs between the present solar irradiance and 1.004~\so for 388 and 560~ppm of \codn, and between 1.004 and 1.012~\so for 200 and 280~ppm of \codn.

\begin{table*}
	\begin{center}
		\centering
		\begin{tabularx}{1.\textwidth}{c  c l c   c   c  l c  c   c   c  l c   c   }
			\cline{1-11}
			&\multicolumn{1}{p{1.5cm}}{\centering Model} &\multicolumn{1}{p{1cm}}{\centering \oz} & \multicolumn{1}{p{1.8cm}}{\centering $TSI$(\wmn)} &\multicolumn{1}{p{2cm}}{\centering [\codn](ppm)} &\multicolumn{1}{p{1cm}}{\centering $T_{S}$(K)}  &\multicolumn{1}{p{1.cm}}{\centering $T_{eff}$(K)} & \multicolumn{1}{p{1cm}}{\centering A} & \multicolumn{1}{p{0.5cm}}{\centering $g_{n}$}& \multicolumn{1}{p{1cm}}{\centering T$_{40}$(K)} &\multicolumn{1}{p{2cm}}{\centering \hmr(\gkgn)} \\\cline{1-11}
			\hhline{===========}
			
			\multicolumn{1}{c}{\centering a}&\multicolumn{1}{c}{ERA}  	      & \multicolumn{1}{c}{yes}& \multicolumn{1}{c}{1361} & 388  & \multicolumn{1}{c}{289.1}  &\multicolumn{1}{c}{255.3}  & 0.294 & 0.392  & \multicolumn{1}{c}{216} & \multicolumn{1}{c}{2.3$\cdot$10$^{-3}$}\\
			\multicolumn{1}{c}{\centering b}&\multicolumn{1}{c}{PlaSim} & \multicolumn{1}{c}{yes}& \multicolumn{1}{c}{1361}  & 388  & \multicolumn{1}{c}{291.0}  & \multicolumn{1}{c}{255.2} & 0.296  & 0.419 & \multicolumn{1}{c}{217} & \multicolumn{1}{c}{7.4$\cdot$10$^{-3}$} \\
			\multicolumn{1}{c}{\centering c}&\multicolumn{1}{c}{LMDZ$^{i}$}   & \multicolumn{1}{c}{no} & \multicolumn{1}{c}{1365} & 376  & \multicolumn{1}{c}{282.8}  &\multicolumn{1}{c}{253.8} & 0.311  & 0.351  & \multicolumn{1}{c}{170} & \multicolumn{1}{c}{1$\cdot$10$^{-5}$}\\
			\multicolumn{1}{c}{\centering d}&\multicolumn{1}{c}{CAM4$^{ii}$}  & \multicolumn{1}{c}{no}& \multicolumn{1}{c}{1361}  & 367  & \multicolumn{1}{c}{289.1}  &\multicolumn{1}{c}{252.0} & 0.329 & 0.423  & \multicolumn{1}{c}{170} & \multicolumn{1}{c}{1$\cdot$10$^{-5}$}\\
			\cline{1-11}		
		\end{tabularx}
		\caption{Comparison of the climate at the present solar irradiance in ERA, \citet{Leconte2013}, \citet{Wolf2015}, and PlaSim data. The ozone ($O_{3}$) concentration, the total solar irradiance ($TSI$), and the $CO_{2}$ concentration are initial conditions. The surface temperature ($T_{S}$), the effective temperature ($T_{eff}$), the Bond albedo (A), and the normalized greenhouse parameter ($g_{n}$) calculations are explained in Section~\ref{mod}. The temperature (T$_{40}$) and the water vapor mixing ratio ($q_{r}$) are both measured at a pressure level of 40~hPa. N\scriptsize{OTE}: \footnotesize {$^{i}$ \citet{Leconte2013}; $^{ii}$ \citet{Wolf2015}.}}\label{tab2}
	\end{center}
\end{table*}

\begin{figure*}[ht]\centering  
	\includegraphics[width=\textwidth]{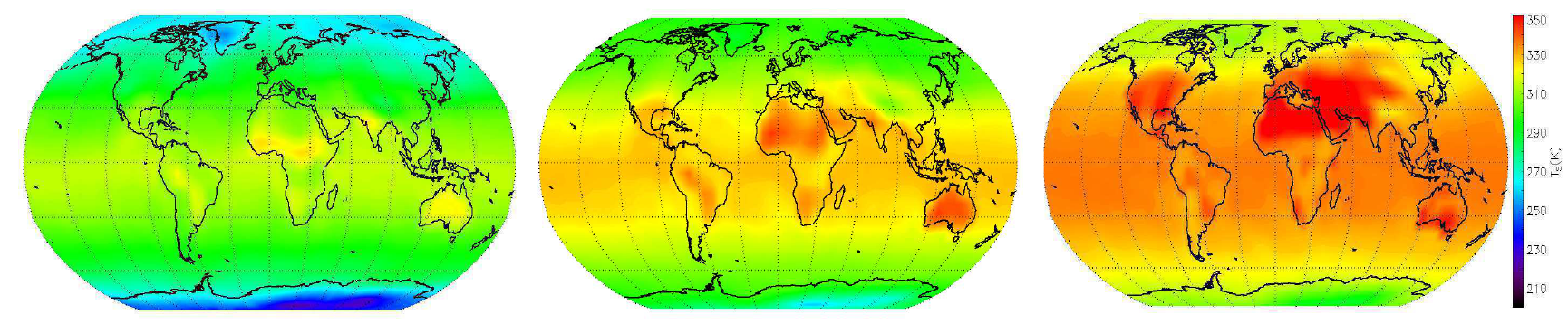}
	\includegraphics[width=\textwidth]{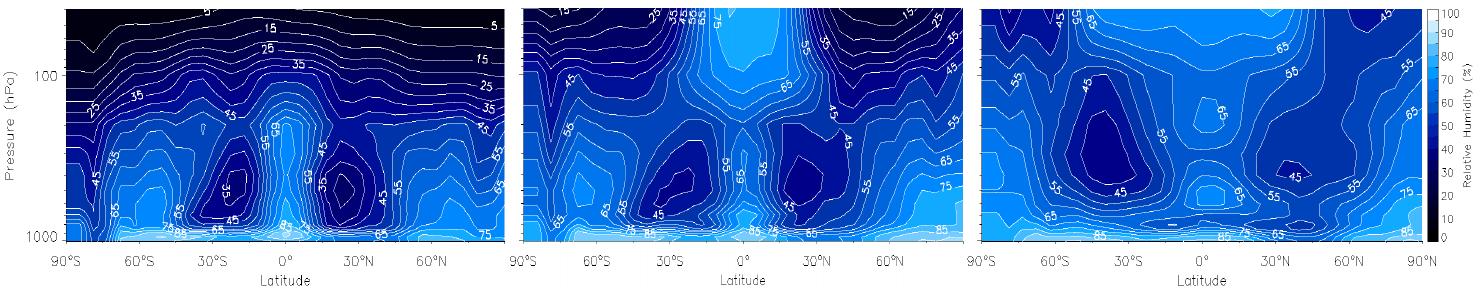}
	\includegraphics[width=\textwidth]{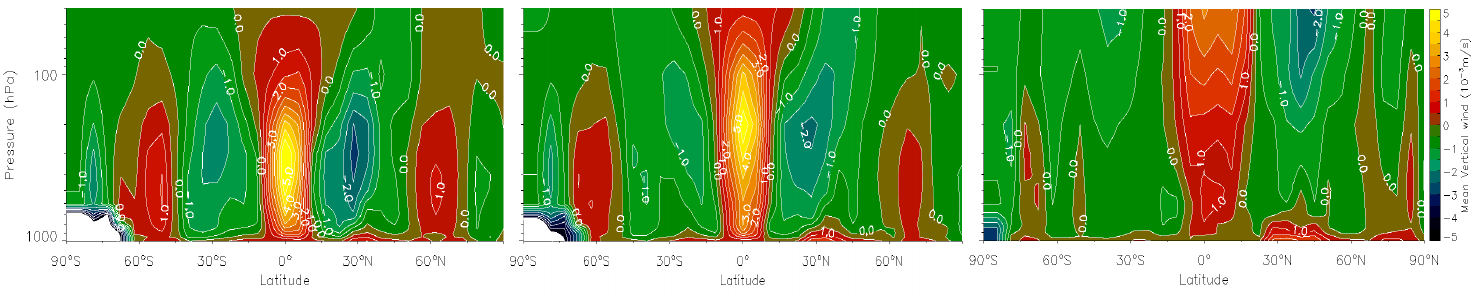}
	\includegraphics[width=\textwidth]{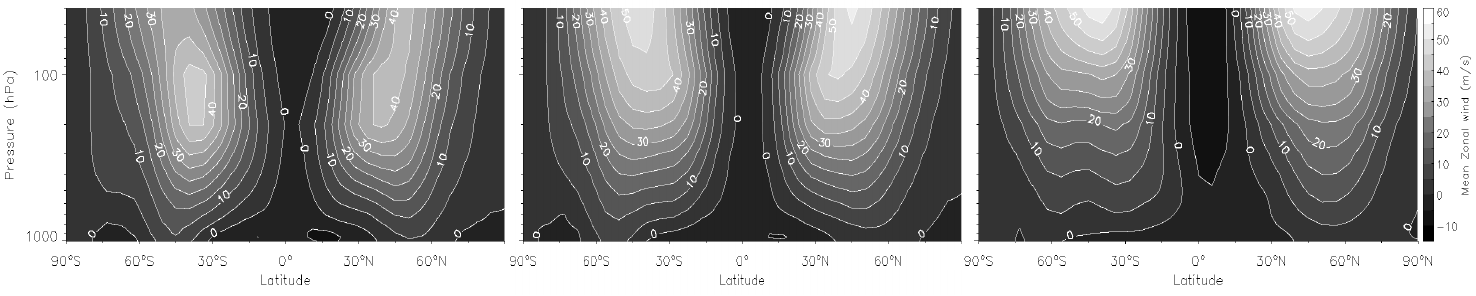}
	\caption{Surface temperature, zonal mean relative humidity (RH), vertical wind, and zonal wind for an atmospheric concentration [\codn]=\,388~ppm, at the present stellar flux (TSI=\son, left), at the state when the polar ice caps melt completely (TSI=1.052~\son, middle), and at the moist greenhouse state (TSI=1.154~\son, right).\label{Fig7}}
\end{figure*}

\begin{figure}[ht]\centering  
	\includegraphics[width=\columnwidth]{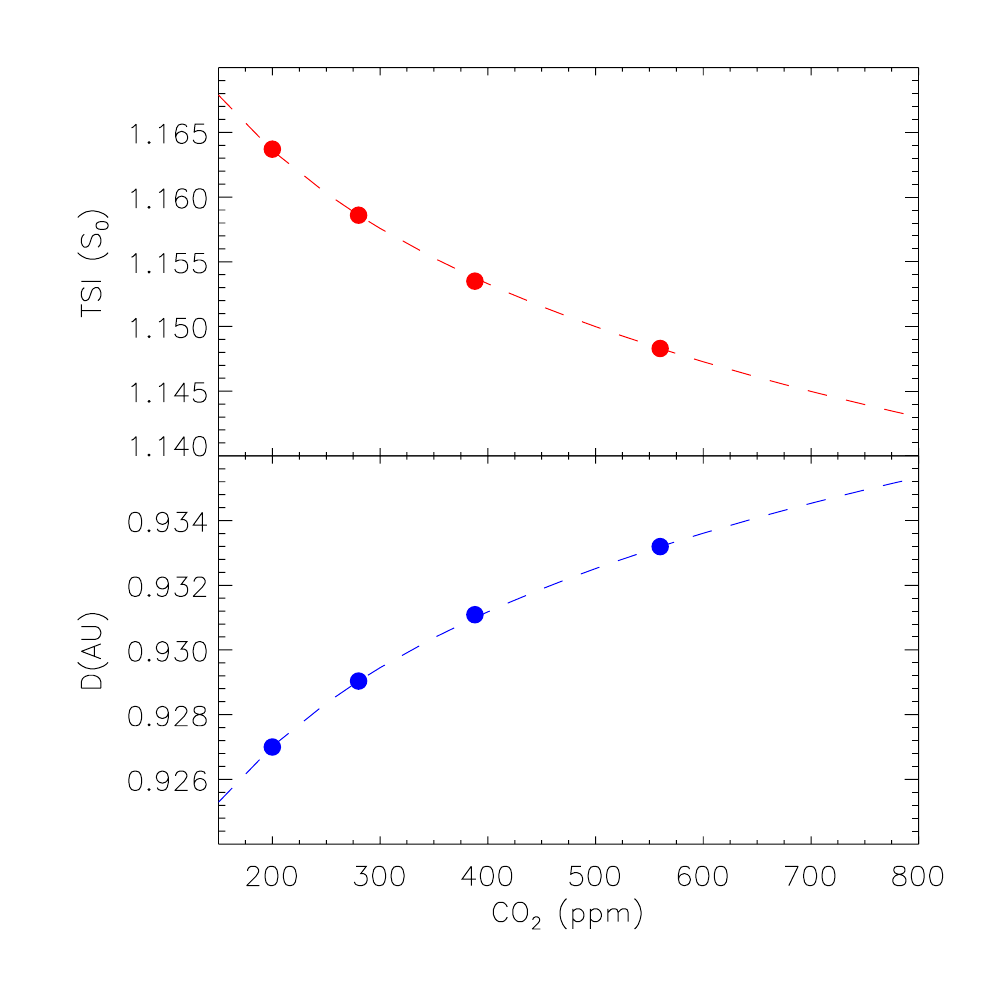}
	\caption{Total solar irradiance (TSI) at the moist greenhouse threshold and
		equivalent orbital distance in our present solar system as a function of the \cod concentration. The TSI has been fitted as $S^{x}_{MGT}=\sigma_{1}\,\ln([CO_{2}]^{x}/[CO_{2}]^{r})+\sigma_{2}$, where  $\sigma_{1}=-1.486\times10^{-2}$~\so and $\sigma_{2}=1.159$~\so (red dash line) and the equivalent orbital distance as $D^{x}_{MGT}=\delta_{1}\,\ln([CO_{2}]^{x}/[CO_{2}]^{0})+\delta_{2}$, where $\delta_{1}=6.00\times10^{-3}$~au and $\delta_{2}=9.29\times10^{-1}$~au (blue dashed line). \label{Fig8}}
\end{figure}

\begin{table}
		\begin{tabularx}{0.5\textwidth}{*{5}{X} }
			\cline{1-5}
			\multicolumn{5}{c}{\small{\emph{Ice Melt}}}\\
			\cline{1-5}
			[${CO}_{2}$](ppm) & \multicolumn{1}{c}{560} & \multicolumn{1}{c}{ 388} & \multicolumn{1}{c}{ 280} & \multicolumn{1}{c}{ 200}\\\cline{1-5}
			$T_{S}^{c}$(K)$^{i}$ & \multicolumn{1}{c}{303.1} & \multicolumn{1}{c}{ 303.0} & \multicolumn{1}{c}{ 303.0} & \multicolumn{1}{c}{ 303.1} \\
			$\epsilon T_{S}^{c}$(K)  & \multicolumn{1}{c}{0.3} & \multicolumn{1}{c}{ 0.3} & \multicolumn{1}{c}{0.4} & \multicolumn{1}{c}{ 0.1} \\
			$S_{c}$($S_{0}$) & \multicolumn{1}{c}{1.0436} & \multicolumn{1}{c}{ 1.0524} & \multicolumn{1}{c}{ 1.0590} & \multicolumn{1}{c}{ 1.0703}  \\
			$\epsilon S_{c}$($S_{0}$) & \multicolumn{1}{c}{ 2$\times$10$^{-4}$}  & \multicolumn{1}{c}{ 2 $\times$10$^{-4}$}  & \multicolumn{1}{c}{ 6 $\times$10$^{-4}$}  & \multicolumn{1}{c}{ 1 $\times$10$^{-4}$}  \\\hhline{=====}
			\multicolumn{5}{c}{\small{\emph{Moist Greenhouse Threshold}}}\\\cline{1-5}
			[${CO}_{2}$](ppm) & \multicolumn{1}{c}{560} & \multicolumn{1}{c}{388} & \multicolumn{1}{c}{280} & \multicolumn{1}{c}{200}\\\cline{1-5}
			$q_{r}^{d}$(\gkgn)$^{ii}$ & \multicolumn{1}{c}{7.6} & \multicolumn{1}{c}{7.5} & \multicolumn{1}{c}{7.5} & \multicolumn{1}{c}{7.5}  \\
			$\epsilon q_{r}^{d}$(\gkgn) & \multicolumn{1}{c}{0.1} & \multicolumn{1}{c}{0.1} & \multicolumn{1}{c}{0.1} & \multicolumn{1}{c}{0.2} \\
			$S_{d}$($S_{0}$) &  \multicolumn{1}{c}{1.1485} & \multicolumn{1}{c}{1.1535} & \multicolumn{1}{c}{1.1586} & \multicolumn{1}{c}{1.1637} \\
			$\epsilon S_{d}$($S_{0}$) & \multicolumn{1}{c}{4$\times$10$^{-4}$} & \multicolumn{1}{c}{4$\times$10$^{-4}$} & \multicolumn{1}{c}{3$\times$10$^{-4}$} & \multicolumn{1}{c}{9$\times$10$^{-4}$} \\
			$T_{st}$(K) & \multicolumn{1}{c}{242.6} & \multicolumn{1}{c}{242.5} & \multicolumn{1}{c}{ 242.8} & \multicolumn{1}{c}{242.9} \\
			$\epsilon T_{st}$(K) & \multicolumn{1}{c}{0.2} & \multicolumn{1}{c}{0.2} & \multicolumn{1}{c}{0.3} & \multicolumn{1}{c}{0.3} \\
			$T_{S}^{d}$(K) & \multicolumn{1}{c}{320.0} & \multicolumn{1}{c}{320.0} & \multicolumn{1}{c}{320.1} & \multicolumn{1}{c}{320.1} \\
			$\epsilon T_{S}^{d}$(K)  & \multicolumn{1}{c}{0.2} & \multicolumn{1}{c}{0.2} & \multicolumn{1}{c}{0.2} & \multicolumn{1}{c}{0.2} \\
			$S_{d}$(\wmn)& \multicolumn{1}{c}{1563} & \multicolumn{1}{c}{1570}& \multicolumn{1}{c}{1577}& \multicolumn{1}{c}{1584}\\
			D(au)& \multicolumn{1}{c}{0.933} & \multicolumn{1}{c}{0.931}& \multicolumn{1}{c}{0.929}& \multicolumn{1}{c}{0.927}\\\cline{1-5}
		\end{tabularx}
		\caption{Polynomial Fittings. Inflection points of the surface temperature ($T_{S}$) series and the water vapor mixing ratio ($q_{r}$) series with respect to the solar irradiance (S, in units of the solar constant $S_{0}$) for different CO$_{2}$ concentrations. ($T_{S}^{c}$,~$S_{c}$) is the inflection point of the polynomial fitting of the surface temperature (Fig.~\ref{Fig1}). (q$_{r}^{d}$,~$S_{d}$) is the inflection point of the polynomial fitting of the water mixing ratio (Fig.~\ref{Fig4}). $T_{S}^{d}$ and $T_{st}$ are the surface temperature and the stratospheric temperature at the inflection point $S_{d}$. D is the equivalent orbital distance in astronomical units (au) for each case.\\
			N\scriptsize{OTE}: \footnotesize{$^{i}$ $T_{S}$(S)=$\sum_{i=0}^{3}\gamma_{i}S^{i}$; 
			$^{ii}$ $q_{r}$(S)=$\sum_{j=0}^{3}\mu_{j}S^{j}$, where $\gamma_{i}$ and $\micron_{j}$ are the polynomial coefficients that depend on the CO$_{2}$ concentration.}}\label{tab3}

\end{table}

\emph{c) The complete melt of the polar ice caps.-- }\label{c}
The second peak of the climate sensitivity occurs when ice and snow have practically disappeared from the planet's surface (Fig.~\ref{Fig1}, state c). This happens for similar conditions of temperature, albedo, and relative humidity (RH) for all the \cod cases tested. As the ice and snow melt, the albedo drops to a minimum value of 0.11, the surface temperature rises to 303~K, and the RH increases to 40\% in the stratosphere.  With more water in the atmosphere, latent heat flux increases (132~\wmn), the temperature difference between the immediate atmosphere and the surface is reduced (Fig.~\ref{Fig6}), and both the sensible heat flux (16.5~\wmn) and the net LW radiation ($\sim$30~\wmn) decrease (Fig.~\ref{Fig3}). The values obtained for these quantities at this state are similar for all the \cod concentrations tested. Although the RH is enhanced, the higher temperatures rise the dew point and both the low cloud fraction (0.12) and the high cloud fraction (0.10) decrease. High clouds form at an upper level (100 hPa) (Fig.~\ref{Fig6}). The Bond albedo and the cloud albedo are reduced by 7\% and by 2\%, respectively, and the emissivity of the atmosphere rises to 0.90. The vertical
temperature gradient increases in the tropics, thus the Hadley cells expand, the intensity peak of the subtropical jet streams move to lower pressure levels, and their speed increases (Fig.~\ref{Fig7}). 

Taking advantage of the large number of simulations performed, we calculate the polynomial fitting of the surface temperature series and its inflection point to determine the climate sensitivity change with a better precision. The results show an inflection point at $T_{S}$$\sim$303~K for all the \cod concentrations tested (Table~\ref{tab3} and Fig.~\ref{Fig1}).

\emph{d) The moist greenhouse effect.-- }\label{d}
The third climate sensitivity peak correlates with the large increase in the humidity of the stratosphere indicating the moist greenhouse threshold (Fig.~\ref{Fig4}, state d). The troposphere is charged with water vapor but it is not completely saturated at this state (RH$\sim$85$\%$ at the surface and RH$\sim$65$\%$ in the stratosphere), in agreement with previous 3D studies \citep{Leconte2013, Wolf2015, Popp2016}. The Hadley cells and the jet streams speed are enhanced with the higher temperatures (Fig.~\ref{Fig7}). The temperature in the surface boundary layer increases (Fig.~\ref{Fig6}), due to the infrared water vapor continuum absorption \citep{London1980}. This phenomenon appears as a temperature inversion in previous studies including a water vapor continuum parameterization \citep{Words2013, Wolf2015, Popp2016}. The sensible heat flux (11~\wmn) and the net LW radiation decrease (8~\wmn) and the latent heat flux increases (180~\wmn). Low and high cloud fractions decrease below 10\% and the emissivity of the atmosphere is enhanced to about 0.99. The values of these quantities are similar for the four \cod concentrations tested.

The fitting curve of the water vapor mixing ratio at 40~hPa shows an inflection point (Fig.~\ref{Fig4}, top chart), indicating the MGT. For 388~ppm of \codn, this occurs at a TSI$\sim$1.154~\so for the present \cod concentration, which corresponds to a distance of 0.930~au in the present Solar System. The water vapor mixing ratio at 40~hPa (q$_{r}\sim7.5$~\gkgn), the stratospheric temperature (T$_{st}\sim243$~K), and the surface temperature (T$_{S}\sim320$~K) have a similar value at this state for all the \cod cases tested (Table~\ref{tab2}, Figs.~\ref{Fig1}, and \ref{Fig4}). In order to take the evolution of the luminosity of a solar type star into account, we use the solar data given by \cite{Bahcall2001}, which have been calibrated with helioseismology measurements. In the case of a pre-industrial \cod concentration level (280~ppm), the moist greenhouse becomes dominant at TSI$\sim$1.159~\so, 50 million years later than with the present
concentration. If the concentration is doubled (560~ppm), the MGT occurs 60 million years earlier than with the present concentration, at TSI$\sim$1.149~\so. In an atmosphere with a
\cod concentration similar to that of some Earth glaciations (200~ppm), the MGT is achieved 100 million years later than with the present concentration, at TSI$\sim$1.164~\son.

Using the relation between the \cod concentration and the radiative forcing (Eq.~\ref{arr}), our results can be fitted by the logarithmic function
\begin{equation}\label{smgt}
S^{x}_{MGT}=\sigma_{1}\,\ln{\Bigg(\frac{[CO_{2}]^{x}}{[CO_{2}]^{r}}\Bigg)} + \sigma_{2}
\end{equation}
where S$^{x}_{MGT}$ is the solar irradiance at the moist greenhouse threshold for a given \cod concentration, [\codn]$^{r}$=280~ppm is the preindustrial \cod concentration, $\sigma_{1}=-1.486\times10^{-2}$~\so, and $\sigma_{2}=S^{r}_{MGT}=1.159$~\so (Table~\ref{tab3} and Fig.~\ref{Fig8}). This function allows us to calculate the MGT for different \cod concentrations. The equivalent distance in our present Solar System can be represented by the function  
\begin{equation}
D^{x}_{MGT}=\delta_{1} \ln{\Bigg(\frac{[CO_{2}]^{x}}{[CO_{2}]^{0}}\Bigg)} + \delta_{2}\label{dmgt}
\end{equation}
where $\delta_{1}=6.00\times10^{-3}$~au and $\delta_{2}=9.29\times10^{-1}$~au. 

Using Equation~\ref{smgt} to obtain $S^{x}_{MGT}$ and $S^{r}_{MGT}$, and supposing that the TOA is in equilibrium in each case (S(1 - A)/4\,=\,$\sigma$\,$T^{4}_{eff}$), we can calculate the difference in the OLR (OLR\,=\,$\sigma$\,$T^{4}_{eff}$) at the MGT between two Earth analog planets with different \cod concentrations by,
\begin{equation}\label{olr}
\Delta(OLR_{MGT})=\mu\,\Bigg(5.35 \,\ln\frac{[CO_{2}]^{x}}{[CO_{2}]^{r}}\Bigg)
\end{equation}
where $\Delta(OLR_{MGT})$ is expressed in \wm and $\mu=\sigma_{1}\,S_{0}(1 - A^{x}_{MGT})/4= -0.72$. Note that the relation inside the parenthesis is Equation~\ref{arr}. However, Equation~\ref{olr} does not represent a radiative forcing but the OLR difference at the MGT between two Earth analogs.

\begin{table}
	\begin{center}
	
		\begin{tabularx}{0.5\textwidth}{*{5}{X}}
			\cline{1-5}
			\multicolumn{5}{c}{\small{\emph{Moist Greenhouse Threshold}}}\\
			\cline{1-5}
			[${CO}_{2}$](ppm) & \multicolumn{1}{c}{560}  & \multicolumn{1}{c}{388} & \multicolumn{1}{c}{280} & \multicolumn{1}{c}{200}\\\cline{1-5}
			$S_{d}$($S_{0}$)$^{i}$ & \multicolumn{1}{c}{1.1485} & \multicolumn{1}{c}{1.1535} & \multicolumn{1}{c}{1.1586} & \multicolumn{1}{c}{1.1637} \\
			$t_{d}$(Gyr)$^{ii}$ & \multicolumn{1}{c}{1.58} & \multicolumn{1}{c}{1.64} & \multicolumn{1}{c}{ 1.69} & \multicolumn{1}{c}{1.74}\\\cline{1-5}
			$\tau_{d}$(Gyr) & \multicolumn{1}{c}{1.44} & \multicolumn{1}{c}{ 1.44} & \multicolumn{1}{c}{1.45} & \multicolumn{1}{c}{1.45} \\\hhline{=====}
			\multicolumn{5}{c}{\small{\emph{Opaque Atmosphere}}}\\\cline{1-5}
			[${CO}_{2}$](ppm) & \multicolumn{1}{c}{560}  & \multicolumn{1}{c}{388} & \multicolumn{1}{c}{280} & \multicolumn{1}{c}{200}\\\cline{1-5}
			$S_{e}$($S_{0}$) & \multicolumn{1}{c}{1.1897} & \multicolumn{1}{c}{1.1897}  & \multicolumn{1}{c}{1.1897}  & \multicolumn{1}{c}{1.1897}\\
			$t_{e}$(Gyr) & \multicolumn{1}{c}{2.00} & \multicolumn{1}{c}{2.00} & \multicolumn{1}{c}{2.00} & \multicolumn{1}{c}{2.00}\\\cline{1-5}
			$\tau_{e}$(Gyr) & \multicolumn{1}{c}{0.83} & \multicolumn{1}{c}{0.84} & \multicolumn{1}{c}{0.86} & \multicolumn{1}{c}{0.88}\\ 
			$\tau_{d,e}$(Gyr) & \multicolumn{1}{c}{1.25} & \multicolumn{1}{c}{1.20} & \multicolumn{1}{c}{1.17} & \multicolumn{1}{c}{1.14}\\
			$\tau_{t}$(Gyr) & \multicolumn{1}{c}{2.83} & \multicolumn{1}{c}{2.84} & \multicolumn{1}{c}{2.86} & \multicolumn{1}{c}{2.88}\\\cline{1-5}
		\end{tabularx}
		\caption{Water loss. [\codn] is the \cod concentration, $S$ is the solar irradiance, $t$ is the time to the future, $\tau_{d}$ is the water lifetime at the triggering of the moist greenhouse effect (state d) calculated following the conditions at that state, $\tau_{e}$ is the water lifetime at the state when the atmosphere becomes opaque (last point of the series), $\tau_{t}$=$t_{e}$+$\tau_{e}$ is the total time from the present to the complete loss of the planet's water, and $\tau_{d,e}$=$\tau_{t}$-$t_{d}$ is the total time from the moist greenhouse threshold to the complete loss of the planet's water. \\
			N\scriptsize{OTE}: \footnotesize{$^{i}$$S(t)=-1.494+1.718\,t-0.469\,t^{2}+0.059\,t^{3}+(-0.003)\,t^{4}$;\\
				$^{ii}\epsilon\,t=0.01$~Gyr.}}\label{tab4}
\end{center}
\end{table}

\emph{e) The opaque atmosphere.-- }\label{e}
The temperature of the planet and the water vapor of the atmosphere continues to increase for larger values of the solar irradiance, and eventually, the opacity and the efficient emissivity ($\epsilon$) of the atmosphere reach their maximum (Fig.~\ref{Fig2}). The sensible heat flux and the net LW radiation at the surface are reduced (Fig.~\ref{Fig3}), due to the intense humidity and the temperature inversion. Most of the energy absorbed by the surface is released in form of latent heat flux. The Simpson-Nakajima limit is reached when $\epsilon=1$. At this point, the OLR depends exclusively on the temperature of the top emitting layer. Our results place this limit at a TSI$\sim1.170$~\so for the present-day \cod concentration. This radiation value is equivalent to an orbital distance of 0.925~au in our present Solar System. In contrast with recent studies \citep{Goldblatt2013}, we obtained an OLR$_{max}\sim310$~\wm (Fig.~\ref{Fig2}), similarly to \citet{Kasting1988}. We have not simulated larger solar forcings, because our model uses a broadband radiative transfer (see Section~\ref{mod}) that it is not adapted to simulate such hot humid states and clouds may form higher than 40~hPa, its top pressure level (Fig.~\ref{Fig6}).

\begin{figure}[ht]\centering  
	\includegraphics[width=\columnwidth]{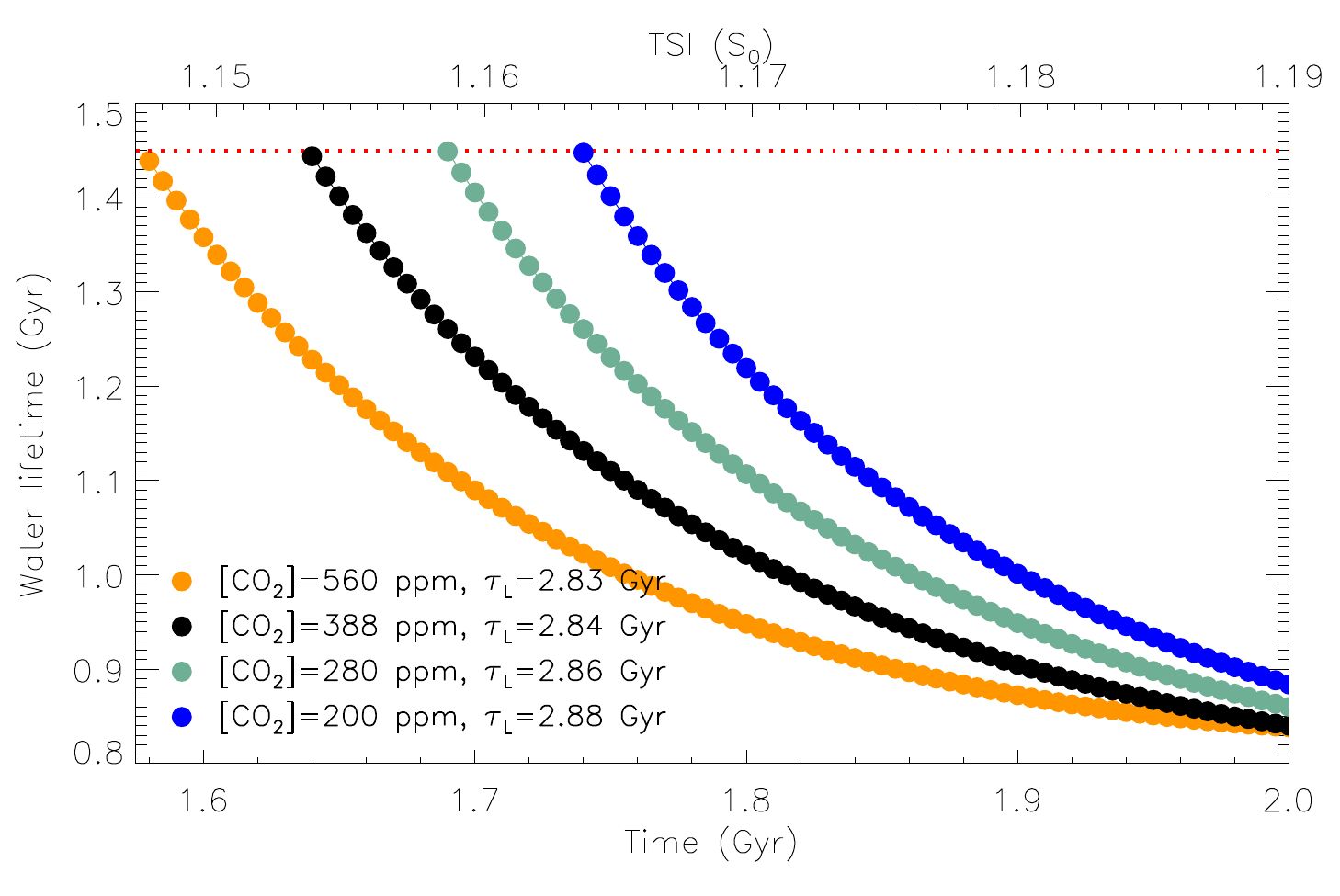}
	\caption{Water lifetime as a function of time and solar irradiance (top axis, in units of the solar constant \son) for four \cod concentrations. The red dotted line corresponds to the moist greenhouse threshold (at this state, the water vapor mixing ratio is about 7.5~\gkg and the surface temperature is 320~K for all the \cod concentrations tested).\label{Fig9}}
\end{figure}

\subsection{The water loss}\label{wl}

Figure~\ref{Fig9} shows the water lifetime of an Earth analog from the MGT to the Simpson-Nakajima limit. The surface temperature of the planet continues to rise with the increase of solar luminosity with time, thus planetary habitability evolves. The water vapor mixing ratio and the escape rate change accordingly and the planet eventually enters into a runaway greenhouse state. We use the solar data given by \cite{Bahcall2001} to account for the evolution of the luminosity with time and we derive the corresponding water vapor mixing ratio value through the polynomial fitting of our model series.

In our simulations of an Earth analog (388 ppm of \codn), the MGT is reached at 1.154~\son. \cite{Bahcall2001} predicts an increase of the solar luminosity to 1.154~\so in 1.64 billion years. Our results show that an Earth analog at the moist greenhouse limit evolves to a Simpson–Nakajima limit, losing the ocean’s water after 1.25 to 1.14 billion years, depending on the \cod levels tested (Fig.~\ref{Fig9} and Table~\ref{tab4}). If the same planet is at the Simpson–Nakajima limit, it loses its water after 0.83 to 0.88 billion years, depending on the \cod concentration. Thus, an Earth analog with the present \cod concentration would enter in a moist greenhouse state and would gradually lose the full water content of the ocean within about 2.84 billion years.

\begin{figure}[ht]\centering  
	\includegraphics[width=\columnwidth]{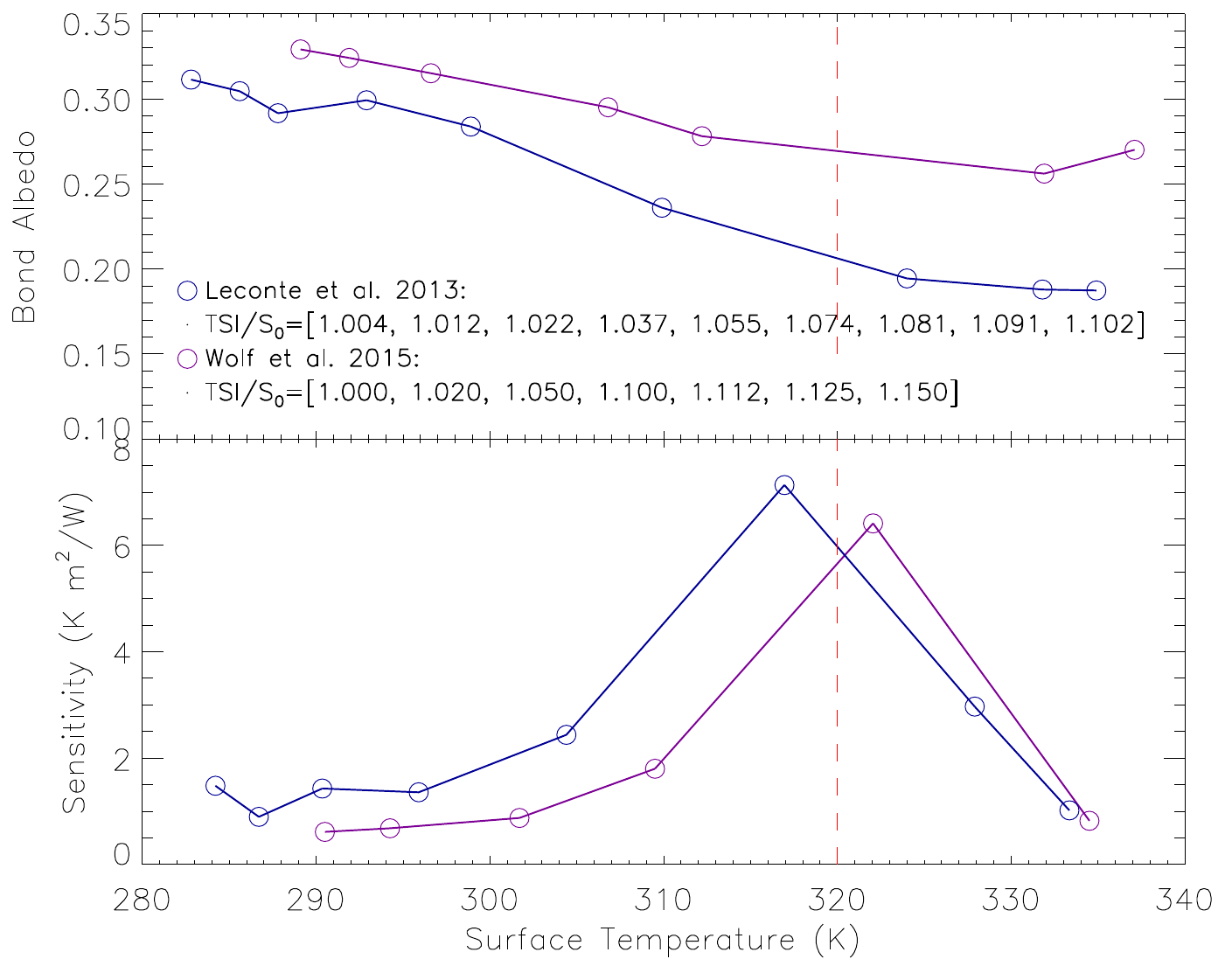}
	\caption{Bond albedo and climate sensitivity as a function of global mean
surface temperature for the experiments in \cite{Leconte2013} (\codn=376~ppm) and in \cite{Wolf2015} (\codn=367~ppm). The solar irradiance at each point is given in units of the solar constant (\son=1361.27~\wmn). The red dashed line indicates the surface temperature standard for the moist greenhouse threshold obtained in this article. \label{Fig10}}
\end{figure}

\section{Discussion}

\subsection{Comparison with previous GCM studies}\label{com}
We compare our results with two previous GCM studies on the greenhouse effect of Earth-like planets (Fig~\ref{Fig10}). \citet{Leconte2013} (hereafter L13) simulates an Earth-like planet with an atmosphere composed by 1~bar of N$_{2}$, 376~ppm of \codn, a variable amount of \ag, and an initial TSI=1365~\wm using a modified version of the LMD Generic GCM (LMDG); \citet{Wolf2015} (heareafter W15) uses a modified version of the Community Atmosphere Model (CAM4), with a similar composition of the atmosphere, 367~ppm of \codn, and an initial TSI=1361.27~\wmn. Both models have a photochemical atmosphere, a correlated-k radiative transfer scheme, implementing HITRAN 2008 and HITRAN 2004 k-coefficients, respectively, and include the water vapor continuum. They both use a mixed layer ocean scheme and a thermodynamic sea-ice model. The parameterization of the cumulus convection is different in each model: LMDG uses a moist convective adjustment scheme \citep{Manabe1965, Forget1998}, PlaSim uses a Kuo-type scheme \citep{Kuo1965, Kuo1974}, and CAM4 uses a mass-flux scheme \citep{Zhang1995}. The last two schemes represent the penetrative cumulus convection and its interaction with the environment, which are important to simulate a proper distribution of the humidity and the clouds, while the convective adjustment does not include these effects. L13 accounts for the non-dilute regime of water by including a numerical scheme to calculate the atmospheric mass redistribution during condensation.

One of the main differences between this study and previous ones is that our simulations include atmospheric ozone. The structure of the atmosphere and the climate of Earth analogs without ozone differ substantially from that of our planet. For this reason, the climate simulations in L13 and W15 for the present solar irradiance differ with respect to ERA data (Table~\ref{tab2} and Fig.~\ref{Fig5}). Their tropopause levels at the present solar irradiance (about 3~hPa in L13 and at 10~hPa in W15) belong to the high stratosphere and the mesosphere in the present-day Earth. They obtain a lower stratospheric temperature (170~K at 40~hPa), and a lower water mixing ratio at that level (\hmr$\sim10^{-5}$~\gkg for both models). Additionally, W15 shows an albedo $\sim$12\% higher and L13 obtains a surface temperature 6~K colder than our planet.

Although these previous GCM studies have similar initial conditions, they show large differences at the same solar irradiance (Table ~\ref{tab2}). For instance, L13 shows a surface temperature of about 285~K at the present solar irradiance, while in W15 is about 289~K. The surface temperature is about 335~K at TSI$\sim$1.102~\so in L13, while W15 shows a value of about 312~K at TSI$\sim$1.100~\so (Figure 10). W15 compared the climate sensitivity of both models, obtaining a peak between [1.074~\son, 1.081~\son] for L13 data, and a peak between [1.112~\son, 1.125~\son] for W15 data. These radiation values correspond to orbital distances about
0.96~au and 0.94~au in the present solar system. Previous studies on the MGT measure qr at the lower stratosphere following the tropopause level, which increases with temperature \citep[e.g.][]{Kasting1993, Kopparapu2013, Kasting2015}. Following this, W15 reported a \hmr$\sim1.5\times10^{-2}$~\gkg at the climate sensitivity peak, which differs with the value of 3~\gkg obtained by \cite{Kasting1993}. A water vapor mixing ratio \hmr$>1$~\gkg is only reached at 1.19~\so in W15. In contrast with these studies, we compare the water vapor mixing ratio of each simulation at same pressure level (40~hPa), since this level has been chosen as a compromise between the photodissociation and the concentration of water in the atmosphere. At the peak of climate sensitivity, the water mixing ratio value at 40~hPa is between [0.2,10]~\gkg in L13 and between [1,10]~\gkg in W15. Both intervals are in agreement with 7.5~\gkgn, the value obtained in our
simulations. 

Our results are consistent with both ERA and W15 data (Fig.~\ref{Fig5}). Due to the effect of atmospheric ozone, our simulations of present-day Earth’s climate obtain a better agreement with respect to ERA data than previous studies on Earth-like planets that do not include ozone. Despite the limitations of PlaSim, the variation of the mean global values
with the surface temperature is similar to W15 data. We want to
remark that the climate evolution of the atmosphere is not taken
into account, since it requires a deeper understanding of the
climate feedback effects and the capability to simulate them.
Therefore, our results do not represent the climate evolution of
a single planet, but the climate of Earth analogs with similar
atmospheric composition and different solar irradiance.

\subsection{The Moist Greenhouse Threshold}
An Earth analog planet with 388~ppm of \cod reaches the
MGT at TSI$\sim$1.154~\son, corresponding to an equivalent orbital
distance of 0.931~au in the present solar system (Table~\ref{tab3}). At
larger concentrations of \codn, less stellar irradiance is needed to
reach the MGT. Both S$_{MGT}$ and the $\Delta(OLR_{MGT})$ (Eqs.~\ref{smgt} and \ref{olr}, respectively) have a logarithmic dependence on the \cod concentration, being consistent with Equation~\ref{arr}. Note, however, that the coefficient values of the functions derived in this article might depend on the complexity of the model used.

Our simulations show that the global mean surface temperature at the MGT is 320~K, independent of the \cod concentrations tested. L13 and W15 did not run simulations near 320~K, but despite the differences at the present Earth’s
state, their data show a climate sensitivity peak between 310~K and 330~K (Fig.~\ref{Fig10}), compatible with the MGT. Therefore, the temperature value proposed in this article is consistent with these two GCM studies.

\subsection{The Water Loss}
Our results of the water lifetime of an Earth analog differ from previous studies, because we identify the MGT by the inflection point of the water vapor mixing ratio series and we use the value of the water mixing ratio given by our model at this point (\hmr$\sim7.5$~\gkgn), instead of using the earlier 1D
model value (\hmr$\sim3$~\gkgn, \citealt{Kasting1993}). In addition,
we use solar data from a more recent solar model (\citealt{Bahcall2001}, instead of \citealt{Gough1981}). As a consequence, the overall water lifetime on an Earth analog is reduced from 4.6 billion years \citep{Kasting1993} (1D) and 3.50 billion years \citep{Wolf2015} (3D) to 2.84 billion years (this paper).
These results do not take in account the evolution of the climate beyond the Simpson–Nakajima limit, which implies a further increase in temperature and a decrease of the water lifetime, and they do not include other factors that may substantially modify them: for instance, it does not take in account the amount of water lost from the beginning of the moist greenhouse to the present conditions of the planet; changes in sea level and salinity, due to the melt of the polar ice caps and the moist greenhouse effect, will modify the evaporation rates and vary the temperature of the planet \citep{Cullum2016}; the recombination of water molecules decreases the hydrogen atoms reaching the top of the atmosphere; the modification of the ocean transport will have an impact on the climate \citep{Knietzsch2015}; the chemical evolution of the atmosphere; the variation of the solar UV radiation will change the photolysis rate of water \citep{Claire2012}, etc. These calculations are highly dependent on the value of the water mixing ratio. Therefore, the improvement of GCMs is essential to understand the role of the processes involved in the loss of Earth’s water and to make estimates for other planets.

\section{Conclusions}
We use PlaSim, an intermediate complexity model, to perform a large number of GCM simulations in order to constrain the conditions of the MGT in Earth analogs. We include the radiative effect of ozone for the first time in GCM studies of Earth analogs. As a consequence, our representation of the current Earth’s climate is in better agreement with ERA data than previous GCM studies of the MGT that do not include ozone. We explore the climate sensitivity to \codn. We identify three states where the planetary climate changes significantly: (i) the state of maximum cloud fraction, (ii) the complete melt of planet’s ice and snow, and (iii) a large increase in the humidity of the stratosphere, corresponding to the MGT. In order to identify the increase in the stratospheric water vapor that characterizes the MGT for the first time in 3D simulations, we calculate the inflection point of the water vapor mixing ratio curve at 40~hPa. Since the evolution of both the stratosphere and the cold trap are not yet well understood, this pressure level represents a compromise between the dissociation and the concentration of water.

Our results show that, on an Earth-like planet with a \cod concentration similar to the present level, the MGT is reached at a TSI of 1.154~\son, corresponding to an equivalent orbital distance of 0.931~au in our solar system, which represents a new value for the inner edge of the Habitable Zone for Earth analogs with ozone, using an intermediate complexity GCM.
The solar incoming radiation should increase to this value in about 1.64 billion years. In agreement with previous GCM studies, the troposphere is not completely saturated at this state in our simulations and there is a temperature increase in the low troposphere due to the water continuum absorption. Our results show that the irradiance at the MGT and the amount of atmospheric \cod follow a logarithmic relation, consistent with the dependence of the \cod radiative forcing with its concentration.

We update previous calculations of the water lifetime on an Earth analog planet by using the value of the water mixing ratio given by our model at the MGT (\hmr$\sim7.5$~\gkgn), instead of using the value earlier obtained by 1D models (\hmr$\sim3$~\gkgn). By using the value of the water mixing ratio given by our model and taking in account the radiative effect of ozone, we
obtain a shorter water lifetime than previous studies, 2.84 billion years for an Earth analog, compared to 3.50 billion years \citep{Wolf2015} and 4.6 billion years \citep{Kasting1993}.

In our simulations, the moist greenhouse effect is initiated by a large increase in the humidity of the stratosphere at a mean surface temperature T$_{S}\sim$320~K, independent of the \cod concentration. Despite the modeling differences, this surface temperature value is consistent with previous GCM studies \citep{Leconte2013, Wolf2015}, suggesting that both the increase in the humidity of the stratosphere and a global mean surface temperature of 320~K might be robust indicators of the MGT in GCM simulations of Earth-like planets. These results should be further assessed using complex GCMs.

\acknowledgments
We thank Eric Wolf and Jérémy Leconte for making
available CAM4 and LMDG data available to us, and for the
valuable discussion. The authors acknowledge support by the
Simons Foundation (SCOL \#290357, Kaltenegger), the Carl
Sagan Institute, and the Centre for the Mathematics of Planet
Earth of the University of Reading.


\hspace{0.25in}
\bibliographystyle{aasjournal}

\begin{thebibliography}{}
\expandafter\ifx\csname natexlab\endcsname\relax\def\natexlab#1{#1}\fi
\providecommand{\url}[1]{\href{#1}{#1}}
\providecommand{\dodoi}[1]{doi:~\href{http://doi.org/#1}{\nolinkurl{#1}}}
\providecommand{\doeprint}[1]{\href{http://ascl.net/#1}{\nolinkurl{http://ascl.net/#1}}}
\providecommand{\doarXiv}[1]{\href{https://arxiv.org/abs/#1}{\nolinkurl{https://arxiv.org/abs/#1}}}
\sloppy

\bibitem[Abe et al.(2011)]{Abe2011} Abe, Y., Abe-Ouchi, A., Sleep, N.~H., \& Zahnle, K.~J. 2011. \asbio, 11, 443. \dodoi{10.1089/ast.2010.0545} 

\bibitem[Arrhenius(1896)]{Arrhenius1896} Arrhenius, S. 1896. London Edinburgh Dublin Phil. Mag. J. Sci., 41, 237. \dodoi{10.1080/14786449608620846} 

\bibitem[Bahcall et al.(2001)]{Bahcall2001} Bahcall, J.~N., Pinsonneault, M.~H., \& Basu, S. 2001. \apj, 555, 990. \dodoi{10.1086/321493}

\bibitem[Berger(1978a)]{Berger1978a} Berger, A.~L. 1978a. \jas, 35, 2362. \dodoi{0.1175/1520-0469(1978)035<2362:LTVODI>2.0.CO;2}

\bibitem[Berger(1978b)]{Berger1978b} Berger, A.~L. 1978b. QuRes, 9, 139. \dodoi{10.1016/0033-5894(78)90064-9}

\bibitem[Bindoff et al.(2013)]{Bindoff2013} Bindoff, N.~L., Stott, P.~A., Achuta Rao, et al. 2013. In Climate Change 2013. Stocker, T.,~F., et al. eds. Cambridge University Press, 867. \dodoi{10.1017/CBO9781107415324.022}

\bibitem[Boer et al.(1984)]{Boer1984} Boer, G. J., McFarlane, N.~A., Laprise, R., et al. 1984. \ato, 22, 397. \dodoi{10.1080/07055900.1984.9649208}

\bibitem[Bordi et al.(2012)]{Bordi2012} Bordi, I., Fraedrich, K., Sutera, A., \& Zhu, X. 2012. ThApC, 109, 253. \dodoi{10.1007/s00704-011-0579-5}

\bibitem[Boschi et al.(2013)]{Boschi2013} Boschi, R., Lucarini, V., \& Pascale, S.\ 2013. \icarus, 226, 1724. \dodoi{10.1016/j.icarus.2013.03.017}

\bibitem[Claire et al.(2012)]{Claire2012} Claire, M.~W., Sheets, J., Cohen, M., et al. 2012. \apj, 757, 95. \dodoi{10.1088/0004-637X/757/1/95} 

\bibitem[Collins et al.(2013)]{Collins2013} Collins, M., Knutti, R., Arblaster, J., et al.  2013. In Climate Change 2013. Stocker, T., ~F., et al. eds. Cambridge University Press, 1029. \url{https://
www.cambridge.org/core/books/climate-change-2013-the-physicalscience-
basis/longterm-climate-change-projections-commitments-andirreversibility-
pages-1029-to-1076/AAAA16E52861380EACB922351006
59F7}

\bibitem[Cullum and Stevens(2016)]{Cullum2016} Cullum, J., Stevens, D.~P. 2016. \pnas, 113, 4278. \dodoi{10.1073/pnas.1522034113}

\bibitem[Eliasen et al.(1970)]{Eliasen1970} Eliasen, E., Machenhauer, B., \& Rasmussen, E. 1970. Report No. 2, Institute for Theoretical Meteorology, Copenhaguen University, Denmark. 

\bibitem[Etminan et al.(2016)]{Etminan2016} Etminan, M., Myhre, G., Highwood, E.~J., Shine, K.~P. 2016. \grl, 43, 12. \dodoi{10.1002/2016GL071930} 

\bibitem[Evans et al.(1998)]{Evans1998} Evans, S.~J., Toumi, R., Harries, J.~E., et al. 1998. \jgr, 103, 8715. \dodoi{10.1029/98JD00265}

\bibitem[Fioletov(2008)]{Fioletov2008} Fioletov; V.~E. 2008. \ato, 46, 39. \dodoi{10.3137/ao.460103}

\bibitem[Forget et al.(1998)]{Forget1998} Forget, F., Hourdin, F., \& Talagrand, O. 1998. \icarus, 131, 302. \dodoi{10.1006/icar.1997.5874}

\bibitem[Forster(2016)]{Forster2016} Forster, P. 2016. AREPS, 44, 85 \dodoi{10.1146/annurev-earth-060614-105156}. 

\bibitem[Forster et al.(2007)]{Forster2007}Forster, P., Ramaswamy, V., Artaxo,  P., et al. \ 2007.\  In: Climate Change 2007. Solomon, S., et al. eds. Cambridge University Press. \url{http://www.ipcc.ch/publications_and_data/ar4/wg1/en/ch2.html}

\bibitem[Fraedrich et al.(2005a)]{Fraedrich2005a} Fraedrich, K., Jansen, H., Kirk, E., et al. 2005. MetZe, 14, 299. \dodoi{10.1127/0941-2948/2005/0043}

\bibitem[Fraedrich et al.(2005b)]{Fraedrich2005b} Fraedrich, K., Jansen, H., Kirk, E., \& Lunkeit, F. 2005. MetZe, 14, 305. \dodoi{10.1127/0941-2948/2005/0044}

\bibitem[Fraedrich \& Lunkeit(2008)]{Fraedrich2008} Fraedrich, K., \& Lunkeit, F. 2008. TellA, 60, 921. \dodoi{10.1111/j.1600-0870.2008.00338.x}

\bibitem[Garcia \& Solomon(1983)]{Garcia1983} Garcia, R.~R., \& Solomon, S. 1983. \jgr, 88, 1379. \dodoi{10.1029/JC088iC02p01379}

\bibitem[Goldblatt et al.(2013)]{Goldblatt2013} Goldblatt, C., Robinson, T.~D., Zahnle, K.~J., \& Crisp, D. 2013. \natgeo, 6, 661. \dodoi{10.1038/ngeo1892}


\bibitem[Gough(1981)]{Gough1981} Gough, D.~O. 1981. SoPh, 74, 21. \dodoi{10.1007/BF00151270}


\bibitem[Green(1964)]{Green1964} Green, A.~E.~S. 1964. ApOpt, 3, 203. \dodoi{10.1364/AO.3.000203}

\bibitem[Hartmann et al.(2013)]{Hartmann2013} Hartmann, D.~L., Klein Tank, A.~M.~G., Rusticucci, M., et al. 2013. In: Climate Change 2013. Stocker, T.~F., et al. eds. Cambridge University Press, 159. \dodoi{10.1017/CBO9781107415324.008}

\bibitem[Hersbach et al.(2015)]{Hersbach2015} Hersbach, H., Peubey, C., Simmons, A., et al. 2015. QJRMS, 141, 2350. \dodoi{10.1002/qj.2528}

\bibitem[Hunten(1973)]{Hunten1973} Hunten, D.~M. 1973. \jas, 30, 1481. \dodoi{10.1175/1520-0469(1973)030<1481:TEOLGF>2.0.CO;2}

\bibitem[Kasting(1988)]{Kasting1988} Kasting, J.~F. 1988. \icarus, 74, 472. \dodoi{10.1016/0019-1035(88)90116-9}

\bibitem[Kasting et al.(2015)]{Kasting2015} Kasting, J.~F., Chen, H., \& Kopparapu, R.~K. 2015. \apj, 813, L3. \dodoi{10.1088/2041-8205/813/1/L3}

\bibitem[Kasting et al.(1984)]{Kasting1984} Kasting, J.~F., Pollack, J.~B., \& Ackerman, T.~P. 1984. \icarus 57, 335. \dodoi{10.1016/0019-1035(84)90122-2}

\bibitem[Kasting et al.(1993)]{Kasting1993} Kasting, J.~F., Whitmire, D.~P., \& Reynolds, R.~T. 1993. \icarus, 101, 108. \dodoi{10.1006/icar.1993.1010}

\bibitem[Katayama(1972)]{Katayama1972}Katayama, A., 1972. Tech. Report, No. 6 (Los Angeles, CA: UCLA Department of Meteorology)

\bibitem[Kilic et al.(2017)]{Kilic2017} Kilic, C., Raible, C.~C., \& Stocker, T.~F. 2017. \apj, 844, 147. \dodoi{10.3847/1538-4357/aa7a03}

\bibitem[Knietzsch et al.(2015)] {Knietzsch2015} Knietzsch, M.-A., Schroder, A., Lucarini, V., \& Lunkeit, F. 2015. ESD, 6, 591. \dodoi{10.5194/esd-6-591-2015}

\bibitem[Kopparapu et al.(2013)]{Kopparapu2013} Kopparapu, R.~K., Ramirez, R., Kasting, et al. 2013. \apj, 765, 131. \dodoi{10.1088/0004-637X/765/2/131}

\bibitem[Kuo(1965)]{Kuo1965} Kuo, H.~L. 1965. \jas, 22, 40. \dodoi{10.1175/1520-0469(1965)022<0040:OFAIOT>2.0.CO;2}

\bibitem[Kuo(1974)]{Kuo1974} Kuo, H.~L. 1974. \jas, 31, 1232. \dodoi{10.1175/1520-0469(1974)031<1232:FSOTPO>2.0.CO;2} 

\bibitem[Lacis and Hansen(1974)]{Lacis1974} Lacis, A.~A., Hansen, J. 1974. \jas, 31, 118. \dodoi{10.1175/1520-0469(1974)031<0118:APFTAO>2.0.CO;2}

\bibitem[Laursen and Eliasen(1989)]{Laursen1989} Laursen, L., Eliasen, E. 1989. TellA, 41, 385. \dodoi{10.3402/tellusa.v41i5.11848} 

\bibitem[Leconte et al.(2013)]{Leconte2013} Leconte, J., Forget, F., Charnay, B., Wordsworth, R., \& Pottier, A. 2013. \nat, 504, 268. \dodoi{10.1038/nature12827}

\bibitem[Linsenmeier et al.(2015)]{Linsenmeier2015} Linsenmeier, M., Pascale, S., \& Lucarini, V. 2015. P\&SS, 105, 43. \dodoi{10.1016/j.pss.2014.11.003}

\bibitem[London(1980)]{London1980} London, J. 1980. Proc. NATO Advanced Study Institute on Atmospheric Ozone. Nicolet, M., \& Aikin, A.~C. eds. (Washington, DC: US DOT), 703. 

\bibitem[Louis(1979)]{Louis1979} Louis, J.-F. 1979. BoLMe, 17, 187. \dodoi{10.1007/BF00117978}

\bibitem[Lucarini et al.(2010a)]{Lucarini2010a} Lucarini, V., Fraedrich, K., \& Lunkeit, F. 2010a. QJRMS, 136. \dodoi{10.1002/qj.543} 

\bibitem[Lucarini et al.(2010b)]{Lucarini2010b} Lucarini, V., Fraedrich, K., Lunkeit, F. 2010b. ACPD, 10, 3699. \dodoi{10.5194/acp-10-9729-2010}

\bibitem[Lucarini et al.(2013)]{Lucarini2013} Lucarini, V., Pascale, S., Boschi, R., Kirk, E., \& Iro, N. 2013. \ana, 334, 576. \dodoi{10.1002/asna.201311903}

\bibitem[Lucarini and Ragone(2011)]{Lucarini2011} Lucarini, V., Ragone, F. 2011. RvGeo, 49, RG1001. \dodoi{10.1029/2009RG000323} 

\bibitem[Lunkeit et al.(2011)]{Lunkeit2011} Lunkeit, F., Fraedrich, K., Jansen, H., et al. 2011. Planet Simulator, Reference Manual. Technical Report, University of Hamburg. \url{https://www.mi.uni-hamburg.de/en/arbeitsgruppen/theoretische-meteorologie/modelle/sources/psusersguide.pdf}

\bibitem[Manabe \& M\"{o}ller(1961)]{Manabe1961}Manabe, S. \& M\"{o}ller, F. 1961. MWRv, 89, 503. \dodoi{10.1175/1520-0493(1961)089<0503:OTREAH>2.0.CO;2}

\bibitem[Manabe et al.(1965)]{Manabe1965} Manabe, S., Smagorinsky, J., \& Strickler, R.~F. 1965. MWRv, 93, 769. \dodoi{10.1175/1520-0493(1965)093<0769:SCOAGC>2.3.CO;2} 

\bibitem[Manabe and Wetherald(1975)]{Manabe1975} Manabe, S., Wetherald, R.~T. 1975. \jas, 32, 3. \dodoi{10.1175/1520-0469(1975)032<0003:TEODTC>2.0.CO;2}

\bibitem[Myhre et al.(1998)]{Myhre1998} Myhre, G., Highwood, E.~J., Shine, K.~P., \& Stordal, F. 1998. GeoRL, 25, 2715. \dodoi{10.1029/98GL01908}

\bibitem[Myhre et al.(2013)]{Myhre2013} Myhre, G., Shindell, D., Br\'{e}on, F.~M., et al. 2013. In: Climate Change 2013. Stocker, T.F., et al. eds. Cambridge University Press, 659. \url{https://www.cambridge.org/core/books/climate-change-2013-the-physicalscience-basis/anthropogenic-and-natural-radiative-forcing/63EB1057C36890FEAA4269F771336D4D}

\bibitem[Nakajima et al.(1992)]{Nakajima1992} Nakajima, S., Hayashi, Y.-Y., \& Abe, Y. 1992. \jas, 49, 2256. \dodoi{10.1175/1520-0469(1992)049<2256:ASOTGE>2.0.CO;2}

\bibitem[Orszag(1970)]{Orszag1970} Orszag, S.~A. 1970. \jas, 27, 890. \dodoi{10.1175/1520-0469(1970)027<0890:TMFTCO>2.0.CO;2}  

\bibitem[Pascale et al.(2011)]{Pascale2011} Pascale, S., Gregory, J.~M., Ambaum, M., \& Tailleux, R. 2011. \cldy, 36, 1189. \dodoi{10.1007/s00382-009-0718-1} 

\bibitem[Poli et al.(2016)]{Poli2016} Poli, P., Hersbach, H., Dee, D. P., et al. 2016. \jcli, 29, 4083. \dodoi{10.1175/JCLI-D-15-0556.1}  


\bibitem[Poli et al.(2013)]{Poli2013}Poli, P., Hersbach, H., Tan, D.~G.~H., et al. 2013. ERA-20C. ERA Report Series 14, 59.  (Reading, England: ECMWF) \url{https://www.ecmwf.int/en/elibrary/
11699-data-assimilation-system-and-initial-performance-.
evaluation-ecmwfpilot-reanalysis}

\bibitem[Popp et al.(2016)]{Popp2016} Popp, M., Schmidt, H., Marotzke, J. 2016.  \natco, 7, 10627. \dodoi{10.1038/ncomms10627}

\bibitem[Ragone et al.(2016)]{Ragone2016} Ragone, F., Lucarini, V., \& Lunkeit, F. 2016. \cldy, 46, 1459. \dodoi{10.1007/s00382-015-2657-3}

\bibitem[Ramirez and Kaltenegger(2014)]{Ramirez2014} Ramirez, R.~M., \& Kaltenegger, L. 2014. \apjl, 797, L25. \dodoi{10.1088/2041-8205/797/2/L25}

\bibitem[Ramirez and Kaltenegger(2016)]{Ramirez2016} Ramirez, R.~M., \& Kaltenegger, L. 2016. \apj, 823, 6. \dodoi{10.3847/0004-637X/823/1/6} 

\bibitem[Rodgers(1967)]{Rodgers1967}Rodgers, C. D., 1967. QJRMS, 93, 43. \dodoi{10.1002/qj.49709339504}

\bibitem[Roeckner et al.(1992)]{Roeckner1992} Roeckner, E., Arpe, K., Rockel, B., et al. 1992. Max-Planck-Institut f{\"u}r Meteorologie, Hamburg. Technical report 93. \url{https://www.mpimet.mpg.de/fileadmin/
publikationen/Reports/MPI-Report_93.pdf}

\bibitem[Sasamori(1968)]{Sasamori1968} Sasamori, T. 1968. JApMe, 7, 721. \dodoi{10.1175/1520-0450(1968)007<0721:TRCCFA>2.0.CO;2}  

\bibitem[Schr{\"o}der and Connon Smith(2008)]{Schroeder2008} Schr{\"o}der, K.-P., Connon Smith, R. 2008. MNRAS, 386, 155. \dodoi{10.1111/j.1365-2966.2008.13022.x}

\bibitem[Simpson(1927)]{Simpson1927} Simpson, G.~C. 1927. QJRMS, 53, 213. \dodoi{10.1002/qj.49705322303}

\bibitem[Slingo and Slingo(1991)]{Slingo1991} Slingo, A., Slingo, J.~M. 1991. \jgr, 96, 15. \dodoi{10.1029/91JD00930}

\bibitem[Stephens(1978)]{Stephens1978} Stephens, G.~L. 1978. \jas, 35, 2123. \dodoi{10.1175/1520-0469(1978)035<2123:RPIEWC>2.0.CO;2}

\bibitem[Stephens et al.(1984)]{Stephens1984} Stephens, G.~L., Ackerman, S., Smith, E.~A. 1984. \jas, 41, 687. \dodoi{10.1175/1520-0469(1984)041<0687:ASPRTI>2.0.CO;2}

\bibitem[Tian et al.(2009)]{Tian2009} Tian, W., Chipperfield, M.~P., L{\"u}, D. 2009. AdAtS, 26, 423. \dodoi{10.1007/s00376-009-0423-3}

\bibitem[Tiedtke(1988)]{Tiedtke1988} Tiedtke, M. \ 1988.\ In Physically-Based Modelling and Simulation of Climate and Climatic Change. Schlesinger M.E., ed. (Dordrecht: Springer), 375. \url{https://www.springer.com/la/book/9789027727886.}

\bibitem[Towe(1981)]{Towe1981} Towe, K.~M. 1981. PreR, 16, 1. \dodoi{10.1016/0301-9268(81)90002-4.}

\bibitem[Trenberth et al.(2009)]{Tren2009} Trenberth, K.~E., Fasullo, J.~T., \& Kiehl, J. 2009. BAMS, 90, 311. \dodoi{10.1029/201609GL037527}

\bibitem[Wolf \& Toon(2015)]{Wolf2015} Wolf, E.~T., \& Toon, O.~B. 2015. \jgrd, 120, 5775. \dodoi{10.1002/2015JD023302} 

\bibitem[Wordsworth and Pierrehumbert(2013)]{Wordsworth2013} Wordsworth, R.~D., \& Pierrehumbert, R.~T. 2013. \apj, 778, 154. \dodoi{10.1088/0004-637X/778/2/154}

\bibitem[Yang et al.(2014)]{Yang2014} Yang, J., Bou{\'e}, G., Fabrycky, D.~C., \& Abbot, D.~S. 2014. \apjl, 787, L2. \dodoi{10.1088/2041-8205/787/1/L2}

\bibitem[Zhang \& MacFarlane(1995)]{Zhang1995} Zhang, G.~J., \& McFarlane N.~A. 1995. \ato, 33, 407. \dodoi{10.1080/07055900.1995.9649539}

\end{thebibliography}

\end{document}